\documentclass[12pt,thmsa]{article}
\usepackage{amssymb}

\usepackage{sw20lart}

\input{tcilatex}
\begin{document}

\title{Stochastic electrodynamics and the interpretatiion of quantum theory}
\author{Emilio Santos}
\maketitle
\tableofcontents

\begin{abstract}
I propose that quantum mechanics is a stochastic theory and quantum
phenomena derive from the existence of real vacuum stochastic fields filling
space. I revisit stochastic electrodynamics (SED), a theory that studies
classical systems of electrically charged particles immersed in an
electromagnetic (zeropoint) radiation field with spectral density
proportional to the cube of the frequency, Planck's constant appearing as
the parameter fixing the scale. Asides from briefly reviewing known results,
I make a detailed comparison between SED and quantum mechanics. Both
theories make the same predictions when the stochastic equations of motion
are of first order in Planck constant, but not in general. I propose that
SED provides a clue for a realistic interpretation of quantum theory.
\end{abstract}

\section{Introduction}

\subsection{Charges immersed in a random (vacuum) radiation field}

The basic assumption in this article is that the quantum vacuum fields are
real stochastic fields. For the sake of clarity let us consider the best
known vacuum field, the electromagnetic zeropoint radiation. The spectrum,
that here I define as the energy per unit volume and unit frequency
interval, is given by eq.$\left( \ref{roZPF}\right) $ below. The parameter
fixing the scale of the field is Planck constant. Therefore it is
interesting to see whether the reality of the vacuum electromagnetic field,
combined with classical physics, allows to explain some phenomena believed
as typically quantal, thus providing a hint for the realistic interpretation
of quantum theory. With that purpose I shall study a restricted domain of
phenomena with a theory defined by:

1) Just one of the interactions of nature, that is electromagnetic, a choice
that we expect should lead to an approximation of quantum electrodynamics
(QED). Actually QED includes two (quantum) vacuum fields, namely
electromagnetic and electron-positron.

2) Nonrelativistic energies. Thus we shall exclude positrons and study only
electrons ignoring spin, or more generally charged particles without
structure, in given electromagnetic fields and/or interacting with other
charges.

3) Planck constant $
\rlap{\protect\rule[1.1ex]{.325em}{.1ex}}h%
$ appears exclusively in the vacuum electromagnetic radiation. Consequently
for the evolution we shall use the laws of classical electrodynamics
throughout, but taking the additional force of the vacuum field into account.

\subsection{Stochastic electrodynamics}

The theory defined with these constraints is already known with the name
stochastic (or random) electrodynamics (SED in the following). It has been
developed by a small number of people during the last sixty years. Actually
SED may be defined in several slightly different forms, for instance as
classical electrodynamics modified by the assumption that the vacuum is not
empty but there is a random electromagnetic field (or zeropoint field, ZPF)
with spectrum eq.$\left( \ref{roZPF}\right) $ filling the whole space. With
the precise definition here proposed SED is an approximation to quantum
electrodynamics to lowest nontrivial order in the Planck constant, the
zeroth order being purely classical electrodynamics. A review of the work
made until 1995 is the book by L. de la Pe\~{n}a and A. M. Cetto\cite{dice}
and new results are included in more recent reviews \cite{dice2}, \cite
{Boyer19}. The application of similar ideas to optics will be reviewed in
Chapter 6.

The origin of SED may be traced back to Walter Nernst, who extended to the
electromagnetic field the zeropoint fluctuations of oscillators assumed by
Planck in his second radiation theory of 1912. Nernst also suggested that
the zeropoint fluctuations might explain some empirical facts, like the
stability of atoms and the chemical bond. The proposal was soon forgotten
due to the success of Bohr\'{}s model of 1913 and the subsequent development
of the (old) quantum theory. Many years later the idea has been put forward
again several times (e. g. by Braffort et al. in 1954\cite{Braffort} and by
Marshall in 1963\cite{Marshall}).

SED studies the motion of charged particles immersed in ZPF, but the back
actions of the charged particles on the ZPF are neglected (indeed the
effects would be of higher order in Planck constant), so that the random
field of free space is used. Assuming that the field is Lorentz invariant
(at not too high frequencies) determines the spectrum, that is the energy
per unit volume and unit frequency interval\cite{Milonni}, \cite{dice}. It
is given by 
\begin{equation}
\rho _{ZPF}\left( \omega \right) =\frac{1}{2\pi ^{2}c^{3}} 
\rlap{\protect\rule[1.1ex]{.325em}{.1ex}}h%
\omega ^{3},  \label{roZPF}
\end{equation}
that corresponds to an average energy $\frac{1}{2} 
\rlap{\protect\rule[1.1ex]{.325em}{.1ex}}h%
\omega $ per normal mode. Planck constant $
\rlap{\protect\rule[1.1ex]{.325em}{.1ex}}h%
$ enters the theory via fixing the scale of the assumed universal random
radiation. Of course the spectrum eq.$\left( \ref{roZPF}\right) $ implies a
divergent energy density and any cutoff would break Lorentz invariance.
However we may assume that it is valid for low enough frequencies, the
behaviour at high frequencies requiring the inclusion of other vacuum fields
and general relativity theory. The spectrum eq.$\left( \ref{roZPF}\right) $
is appropriate for systems at zero Kelvin, but SED may be also studied at a
finite tempereture, where we should add to eq. $\left( \ref{roZPF}\right) $
the thermal Planck spectrum. In addition SED may provide an interpretation
of phenomena where the free spectrum is modified by boundary conditions
derived from macroscopic bodies, but the average energy $\frac{1}{2} 
\rlap{\protect\rule[1.1ex]{.325em}{.1ex}}h%
\omega $ per normal mode still holds true. These phenomena will be revisited
in Chapter 6.

The SED\ study of some simple systems provides an intuitive picture of
several phenomena usually considered as purely quantum, like \textit{the
stability of the classical (Rutherford) atom, Heisenberg uncertainty
relations, entanglement, specific heats of solids, behaviour of atoms in
cavities, etc.} For this reason I propose that SED may be considered as a
clue in the search for a realistic interpretation of quantum theory.

\subsection{Scope of stochastic electrodynamics}

Not all predictions that have been claimed to follow from SED derive from
the theory as defined above. In some cases additional assumptions are
introduced in order to agree with the quantum predictions. In this form most
of nonrelativistic quantum mechanics might be derived from SED\cite{dice2}.
However with extra assumptions not resting upon deep arguments the physical
bases of the theory become unclear and a realistic interpretation
problematic.

In this paper we will study SED strictly as defined in Section 1.1. With
that definition there are many examples where SED predicts results in
contradiction with quantum mechanics and with experiments, as discussed in
Section 6 below. In particular SED deals only with charged particles whilst
QM laws are valid for both charged and neutral particles. It has been
claimed that the restriction may be avoided taking into account that neutral
particles may contain charged parts (e. g. the neutron possesses a magnetic
moment). I think this is flawed, the application to those neutral particles
might be valid in order to explain the stationary equilibrium state, which
is effectively defined over an infinite time and it results independent on
the total charge, as seen for instance in eqs.$\left( \ref{Wx}\right) $ and $%
\left( \ref{Wv}\right) $ below. However this is not the case for time
dependent properties like eq.$\left( \ref{general}\right) $ where the value
of the charge is relevant.

With the definition of Section 1.1 SED is an approximation to QED at the
lowest nontrivial order in Planck constant. A different approach is to
consider that SED is ``the closest classical approximation to quantum
theory'' \cite{Boyer19}. This suggests that there are two different
theories, namely classical and quantum, but it is assumed that the validity
of classical theory may be extended if we include the hypothesis of a
radiation field with a Lorentz invariant spectrum in free space. It seems
that this approach would increase, rather than solve, the problem of the
``infamous boundary'' between classical and quantum theories. Indeed in the
standard wisdom (i. e. ignoring SED) the boundary is defined (roughly) by
the relative value of Planck constant as compared with the typical magnitude
of the action variable for the system. That is, classical theories are an
approximation of quantum theories when Planck constant may be negleted. An
argument for the need of distinguishing between quantum theory and
generalizations of SED is the fact that there are peculiarities of quantum
theory, like discrete spectra, that cannot be achieved by generalized SED.
However there are numerical calculation providing hints that this problem
might be solved, as we will comment on Section 2.5. In any case a
generalization of SED (or equivalently a realistic interpretation of quantum
theory) needs to agree with quantum theory \textit{exclusively} in the
predictions of results of actual experiments, either performed or at least
possible. But it is not required that predictions for ideal (not realistic)
experiments or for unobservable facts should agree.

In this book it is supported the view that the whole of quantum theory
should admit a realistic (classical-like) interpretation. That
interpretation might be obtained via a generalization of SED taking into
account not only the effect of the electromagnetic field on the motion of
charged particles, but also the back action of the particles on the field
and also all other vacuum fields, including metric fluctuations of
spacetime. Attempts in that direction have been made elsewhere\cite{metric}.
In summary SED may be taken as an approximation to quantum electrodynanics
in some limited domain. In particular when the equations of motion are
linear.

\subsection{Plan of the article}

In the following a short review of SED is presented and the analogies and
differences between SED and nonrelativistic quantum mechanics (QM in the
following) for some simple systems are studied. Most of the results have
been reviewed in more detail elsewhere\cite{dice}. The novelty here is a
more careful comparison of SED with QM and the emphasis on those quantum
phenomena that might be better understood via the analogy with the picture
provided by SED.

In the second and third sections the harmonic oscillator is revisited, with
an application to oscillators in several dimensions in section 4. In
sections 5 and 6 SED is applied to other linear systems, namely the free
particle and the particle in a homogeneous magnetic field. Section 7 is
devoted to the application of SED to some nonlinear systems, showing that in
this case some disagreements with QM and with the experiments usualy appear.
Section 8 presents the conclusions. This chapter includes many calculations
and, in order that the reader does not loss the essential points, I will
write in italics the relevant aspects for the comparison between SED and QM.

\section{The harmonic oscillator. Stationary state}

\subsection{Equation of motion}

The harmonic oscillator in one dimension is the most simple system to be
treated within SED (the free particle requires a more careful study in order
to avoid divergences). It is not strange that it was the first system
studied. In this and the following sections we revisit a well known
treatment of the oscillator in SED\cite{S74},\cite{dice}, but the study of
the aspects that may provide a clue for the interpretation of quantum
mechanics is original.

If a charged particle moves in one dimension in a potential well and it is
also immersed in electromagnetic noise, it may arrive at a dynamical
equilibrium between absorption and emission of radiation. In order to study
the equilibrium I shall write the differential equation for the
one-dimensional motion of the particle in the non-relativistic
approximation. The passage to more dimensions is straightforward. We will
neglect magnetic effects of the ZPF and the dependence of the field on the
position coordinate, which corresponds to the common electric dipole
approximation, plausible in a non-relativistic treatment. Thus the
differential equation of motion of the particle in a harmonic oscillator
potential is 
\begin{equation}
m\stackrel{..}{x}=-m\omega _{0}^{2}x+m\tau \stackrel{...}{x}+eE\left(
t\right) ,  \label{ode}
\end{equation}
where $m(e)$ is the particle mass (charge) and $E\left( t\right) $ is the $x$
component of the electric field of the radiation (the zeropoint field, ZPF).
The equation of the mechanical (classical) oscillator is modified by the two
latter terms. The second term on the right side of eq.$\left( \ref{ode}%
\right) ,$ is the damping force due to emission of radiation. It should
appear also in the classical electrodynamical treatment. Only the third term
is specific of SED because it involves Planck constant (it is of order O$%
\left( 
\rlap{\protect\rule[1.1ex]{.325em}{.1ex}}h%
^{1/2}\right) ).$ The parameter $\tau $ given by 
\begin{equation}
\tau =\frac{2e^{2}}{3mc^{3}}\Rightarrow \tau \omega _{0}=\frac{2}{3}\frac{%
e^{2}}{
\rlap{\protect\rule[1.1ex]{.325em}{.1ex}}h%
c}\frac{
\rlap{\protect\rule[1.1ex]{.325em}{.1ex}}h%
\omega _{0}}{mc^{2}}<<1.  \label{gamma}
\end{equation}
so that the dimensionless quantity $\tau \omega _{0}$ is very small, it
being the product of two small numbers namely the fine structure constant, $%
\alpha \equiv e^{2}/
\rlap{\protect\rule[1.1ex]{.325em}{.1ex}}h%
c\sim 1/137,$ and the nonrelativistic ratio $
\rlap{\protect\rule[1.1ex]{.325em}{.1ex}}h%
\omega _{0}/mc^{2}\simeq $ $v^{2}/c^{2}<<1.$ Thus the two latter terms of eq.%
$\left( \ref{ode}\right) $ may be taken as small, which allows some useful
approximations. Eq.$\left( \ref{ode}\right) $ is a stochastic differential
equation of Langevin\'{}s type with coloured (non-white) noise. It has been
named Braffort-Marshall equation by the early workers on SED\cite{Braffort} ,%
\cite{Marshall}. Solving an equation of this kind usually means finding the
evolution of the probability distribution of the relevant quantities as a
function of time, starting from given initial conditions. When the time goes
to infinity the probability distributions become independent of the initial
conditions, giving rise to the stationary or equilibrium distribution.

\subsection{Average values of the potential and kinetic energies}

Several solutions of the eq.$\left( \ref{ode}\right) $ have been published%
\cite{S74}, \cite{dice}. The most simple is the stationary solution, which
may be found by Fourier transform of eq.$\left( \ref{ode}\right) $ as
follows. Firstly we define the Fourier transform of the stationary process $%
E(t)$ in a finite time interval by 
\begin{equation}
\widetilde{E}\left( \omega ,T\right) \equiv \frac{1}{\sqrt{4\pi T}}%
\int_{-T}^{T}E(t)\exp \left( -i\omega t\right) dt.  \label{spectrum0}
\end{equation}
Hence it may be shown that$\left| \widetilde{E}\left( \omega ,T\right)
\right| ^{2}/8\pi $ is the mean (in the time interval $\left( -T,T\right) )$
energy density per unit frequency interval associated to one electric field
component. Thus the total energy density per unit frequency interval, $\rho
\left( \omega \right) $ eq.$\left( \ref{roZPF}\right) ,$ should be $6$ times
that quantity (6 because in the ZPF there are 3 components of the electric
field and another 3 of the magnetic field all contributing equally on the
average). Consequently we define the spectral density, $S_{E}\left( \omega
\right) ,$ of the field $E(t)$ as follows 
\begin{equation}
S_{E}\left( \omega \right) \equiv \lim_{T\rightarrow \infty }\left| 
\widetilde{E}\left( \omega ,T\right) \right| ^{2}=\frac{4\pi }{3}\rho \left(
\omega \right) =\frac{2}{3\pi c^{3}}
\rlap{\protect\rule[1.1ex]{.325em}{.1ex}}h%
\omega ^{3},\smallskip  \label{Espectrum}
\end{equation}
the equality giving the relation between the spectral density and the energy
density of the ZPF, eq.$\left( \ref{roZPF}\right) .$ For short the spectral
density will be named spectrum in the following.

A Fourier transform similar to eq.$\left( \ref{spectrum0}\right) $ of all
terms of eq.$\left( \ref{ode}\right) $ provides a relation between the
spectrum of the field component and the spectrum of the coordinate, $x(t)$,
namely 
\begin{equation}
m(\omega _{0}^{2}-\omega ^{2}+i\tau \omega ^{3})\widetilde{x}\left( \omega
\right) =e\widetilde{E}\left( \omega \right) ,  \label{Fourier}
\end{equation}
where $\widetilde{x}\left( \omega \right) $ and $\widetilde{E}\left( \omega
\right) $ are the Fourier transforms of $x(t)$ and $E(t)$ respectively.
Hence the spectrum of $x\left( t\right) $ is easily got in terms of the
spectrum of $E\left( t\right) $ that is 
\begin{equation}
S_{x}\left( \omega \right) =\lim_{T\rightarrow \infty }\left| \widetilde{E}%
\left( \omega ,T\right) \right| ^{2}=\frac{3c^{3}\tau }{2m\left[ \left(
\omega _{0}^{2}-\omega ^{2}\right) ^{2}+\tau ^{2}\omega ^{6}\right] }%
S_{E}\left( \omega \right) ,  \label{spectrum}
\end{equation}
whence we obtain, taking eq.$\left( \ref{Espectrum}\right) $ into account, 
\begin{equation}
S_{x}\left( \omega \right) =\frac{
\rlap{\protect\rule[1.1ex]{.325em}{.1ex}}h%
\tau \omega ^{3}}{\pi m\left[ \left( \omega _{0}^{2}-\omega ^{2}\right)
^{2}+\tau ^{2}\omega ^{6}\right] }.  \label{oscilspectrum}
\end{equation}

From the spectrum it is trivial to get the quadratic means of the relevant
variables namely 
\begin{equation}
\left\langle x^{2}\right\rangle =\int_{0}^{\infty }S_{x}\left( \omega
\right) d\omega ,\left\langle v^{2}\right\rangle =\int_{0}^{\infty }\omega
^{2}S_{x}\left( \omega \right) d\omega ,\smallskip  \label{mean}
\end{equation}
where $\left\langle {}\right\rangle $ means time average, and the quantities
in eq.$\left( \ref{mean}\right) $ are the coordinate of the oscillator and
its velocity, respectively. The spectrum of the velocity is $\omega ^{2}$
times the spectrum of the coordinate because the time derivative leads to
multiplication of the Fourier transform times $i\omega $. In our treatment
of stationary states in SED an ergodic hypothesis is made, that is ensemble
averages are assumed equal to time averages for the stationary stochastic
processes involved.

Calculating the integral of $S_{x}\left( \omega \right) $ is lengthy but it
becomes trivial in the limit $\tau \rightarrow 0$ where the integrand is
highly peaked at $\omega \simeq \omega _{0}.$ If $\tau $ is small the
contribution to the integral comes only from values of $\omega $ close to $%
\omega _{0}$ and we may put $\omega \rightarrow \omega _{0},$ except in the
difference $\omega -\omega _{0}$, and then to extend the integral to the
whole real line. With this substitution the integrand becomes a Dirac\'{}s
delta in the limit $\tau \rightarrow 0$ and the integral becomes trivial$,$
that is

\begin{eqnarray}
\left\langle x^{2}\right\rangle &=&\int_{0}^{\infty }S_{x}\left( \omega
\right) d\omega \simeq \int_{-\infty }^{\infty }\frac{%
\rlap{\protect\rule[1.1ex]{.325em}{.1ex}}h%
\tau \omega _{0}^{3}}{\pi m\left[ 4\omega _{0}^{2}\left( \omega _{0}-\omega
\right) ^{2}+\tau ^{2}\omega _{0}^{6}\right] }d\omega  \nonumber \\
&\simeq &\frac{
\rlap{\protect\rule[1.1ex]{.325em}{.1ex}}h%
}{2m\omega _{0}}\int_{-\infty }^{\infty }\delta \left( \omega -\omega
_{0}\right) d\omega =\frac{
\rlap{\protect\rule[1.1ex]{.325em}{.1ex}}h%
}{2m\omega _{0}},  \label{2.3}
\end{eqnarray}
whence the mean potential energy is 
\[
\left\langle V\right\rangle =\frac{1}{2}m\omega _{0}^{2}\left\langle
x^{2}\right\rangle =\frac{1}{4}
\rlap{\protect\rule[1.1ex]{.325em}{.1ex}}h%
\omega _{0}. 
\]
The contribution of the high frequencies, $\left\langle x^{2}\right\rangle
_{hf},$ may be approximated by the integral of the spectrum eq.$\left( \ref
{oscilspectrum}\right) $ with zero substituted for $\omega _{0}.$ However,
in order to exclude the low frequency part, calculated in eq.$\left( \ref
{2.3}\right) ,$ we shall put $2\omega _{0}$ as lower limit of the integral,
that is 
\begin{eqnarray}
\left\langle x^{2}\right\rangle _{hf} &\simeq &\int_{2\omega _{0}}^{\infty }%
\frac{
\rlap{\protect\rule[1.1ex]{.325em}{.1ex}}h%
\tau \omega ^{3}}{\pi m\left[ \omega ^{4}+\tau ^{2}\omega ^{6}\right] }%
d\omega  \nonumber \\
&\simeq &-\frac{
\rlap{\protect\rule[1.1ex]{.325em}{.1ex}}h%
\tau }{\pi m}\log \left( \tau \omega _{0}\right) ,  \label{2.3a}
\end{eqnarray}
which is positive (see eq.$\left( \ref{gamma}\right) ).$ We see that the
result depends but slightly on the lower limit of the integral (provided it
is of order $2\omega _{0}).$

A similar procedure might be used for the quadratic mean velocity, by
performing the integral of the velocity spectrum. However that integral is
divergent and we shall assume that there is some frequency cut-off, $\omega
_{c}$. The result of the integral is the sum of two terms. One of them comes
from frequencies near $\omega _{0}$ and it is independent of the cut-off in
the limit $\tau \rightarrow 0$ giving 
\begin{equation}
\left\langle v^{2}\right\rangle =\int_{0}^{\omega _{c}}\omega
^{2}S_{x}\left( \omega \right) d\omega \simeq \frac{%
\rlap{\protect\rule[1.1ex]{.325em}{.1ex}}h%
\omega _{0}}{2m}\Rightarrow \frac{1}{2}m\left\langle v^{2}\right\rangle =%
\frac{1}{4}
\rlap{\protect\rule[1.1ex]{.325em}{.1ex}}h%
\omega _{0}.  \label{2.4}
\end{equation}
The other term comes from the high frequency region and it is divergent when
the cut-off goes to infinity. It may be approximated as in the case of $%
\left\langle x^{2}\right\rangle ,$ although here we may put zero as lower
limit of the integral, that is 
\begin{equation}
\left\langle v^{2}\right\rangle _{hf}\simeq \int_{0}^{\omega _{c}}\frac{%
\rlap{\protect\rule[1.1ex]{.325em}{.1ex}}h%
\tau \omega ^{5}}{\pi m\left[ \omega ^{4}+\tau ^{2}\omega ^{6}\right] }%
d\omega =\frac{
\rlap{\protect\rule[1.1ex]{.325em}{.1ex}}h%
}{2\pi m\tau }\log \left( 1+\tau ^{2}\omega _{c}^{2}\right) .  \label{vhf}
\end{equation}
However that term is not very relevant because for those frequencies the
non-relativistic approximation breaks down (see below the discussion of the
velocity dispersion in the free particle case). Adding eqs.$\left( \ref{2.3}%
\right) $ and $\left( \ref{2.4}\right) $ gives the total mean energy to
zeroth order in the small quantity $\tau \omega _{0}$, namely 
\begin{equation}
\left\langle U\right\rangle =\left\langle \frac{1}{2}m\omega _{0}^{2}x^{2}+%
\frac{1}{2}mv^{2}\right\rangle =\frac{1}{2}
\rlap{\protect\rule[1.1ex]{.325em}{.1ex}}h%
\omega _{0}.  \label{energy}
\end{equation}

An alternative definition of the energy is possible in terms of the
canonical momentum, $p,$ which avoids problems of divergence. The momentum
is defined by 
\begin{equation}
p\equiv mv-\frac{e}{c}A\mathbf{,}U\equiv \frac{p^{2}}{2m}+\frac{1}{2}m\omega
_{0}^{2}x^{2}.  \label{momentum}
\end{equation}
Now we take into account that the potential vector, whose $x$ component we
label $A,$ contains two parts one coming from the ZPF and the other one from
the particle self-field, the latter producing the radiation reaction. These
two terms give rise to the latter two terms of eq.$\left( \ref{ode}\right) .$
Taking this relation into account it is straightforward to get the spectrum
of the canonical momentum, that is 
\begin{eqnarray}
\frac{d}{dt}p &=&-m\omega _{0}^{2}x\Rightarrow S_{p}\left( \omega \right) =%
\frac{m^{2}\omega _{0}^{4}}{\omega ^{2}}S_{x}\left( \omega \right)
\label{canonmom} \\
&=&\frac{
\rlap{\protect\rule[1.1ex]{.325em}{.1ex}}h%
m\tau \omega _{0}^{4}\omega }{\pi \left[ \left( \omega _{0}^{2}-\omega
^{2}\right) ^{2}+\tau ^{2}\omega ^{6}\right] }.  \nonumber
\end{eqnarray}
Hence we get 
\begin{equation}
\left\langle p^{2}\right\rangle =m^{2}\omega _{0}^{4}\int_{0}^{\infty
}\omega ^{-2}S_{x}\left( \omega \right) d\omega =\frac{m 
\rlap{\protect\rule[1.1ex]{.325em}{.1ex}}h%
\omega _{0}}{2}\Rightarrow \frac{\left\langle p^{2}\right\rangle }{2m}=\frac{%
1}{4}
\rlap{\protect\rule[1.1ex]{.325em}{.1ex}}h%
\omega _{0},  \label{2.4a}
\end{equation}
in the limit $\tau \rightarrow 0.$ We see that the energy defined from the
velocity is divergent (a cut-off was needed), whilst the one derived from
the canonical momentum is finite. Thus the use of the canonical momentum in
the definition of the energy seems more convenient. We may expect that in a
more correct relativistic treatment the former would be also convergent and
not too different from the latter.

\subsection{Probability distributions of position, momentum and energy}

In order to fully define the stationary state of the oscillator immersed in
ZPF it is necessary to get the probability distributions, not just the mean
values. Before doing that we need to clarify the meaning of the probability
distributions involved. Up to now we have considered averages over infinite
time intervals, see eq.$\left( \ref{Espectrum}\right) .$ However we assume
that the time dependent quantities are stochastic processes, that is
probability distributions of functions of time. Thus we should write $%
x(t,\lambda )$ (as is standard it the mathematical theory of stochastic
processes) rather than just $x(t)$, where $\lambda \in \Lambda $ and there
is a probability distribution on the set $\Lambda .$ For a fixed value of $t$
this provides a probabiltity distribution of the random variable $x(t)$. We
assume that the probability distribution of each component, $E(t,\lambda )$,
of the ZPF (in free space) is Gaussian with zero mean and also that it is a
stationary ergodic process, that is any time average (over an infinite time
interval) equals the ensemble average over the probability distribution of $%
\Lambda $ at any single time.

Eq.$\left( \ref{ode}\right) $ is linear, whence the Gaussian character of $%
E(t,\lambda )$ gives rise to Gaussian distributions (with zero mean) for
both positions and velocities. Thus eq.$\left( \ref{2,3}\right) $ fixes
completely the normalized probability distribution of the positions to be

\begin{equation}
W\left( x\right) dx=\sqrt{\frac{m\omega _{0}}{\pi 
\rlap{\protect\rule[1.1ex]{.325em}{.1ex}}h%
}}\exp \left[ -\frac{m\omega _{0}x^{2}}{2
\rlap{\protect\rule[1.1ex]{.325em}{.1ex}}h%
}\right] dx.  \label{Wx}
\end{equation}
Similarly eq.$\left( \ref{2.4a}\right) $ fixes the distribution of momenta,
that is 
\begin{equation}
W\left( p\right) dp=\sqrt{\frac{m}{\pi 
\rlap{\protect\rule[1.1ex]{.325em}{.1ex}}h%
\omega _{0}}}\exp \left[ -\frac{p^{2}}{2
\rlap{\protect\rule[1.1ex]{.325em}{.1ex}}h%
m\omega _{0}}\right] dp,  \label{Wv}
\end{equation}
which is also normalized. The distribution of velocities is similar to that
one, with $mv$ substituted for $p$ (modulo ignoring the part due to high
frequencies).

In order to get the distribution of energy, $U$, to lowest order in Planck
constant $
\rlap{\protect\rule[1.1ex]{.325em}{.1ex}}h%
$ we take into account that, as eqs.$\left( \ref{Wx}\right) $ and $\left( 
\ref{Wv}\right) $ already contain $
\rlap{\protect\rule[1.1ex]{.325em}{.1ex}}h%
$, the relation between $x,v$ and $U$ should be written to zeroth order in $
\rlap{\protect\rule[1.1ex]{.325em}{.1ex}}h%
,$ that is using the classical relation. Then we get the following
exponential distribution of energies, $U$, 
\begin{eqnarray}
W\left( U\right) dU &=&\int W\left( x\right) dx\int W\left( p\right)
dp\delta \left( U-\frac{1}{2}m\omega _{0}^{2}x^{2}-\frac{1}{2m}p^{2}\right)
dU  \nonumber \\
&=&\frac{2}{
\rlap{\protect\rule[1.1ex]{.325em}{.1ex}}h%
\omega _{0}}\exp \left( -\frac{2U}{
\rlap{\protect\rule[1.1ex]{.325em}{.1ex}}h%
\omega _{0}}\right) dU,U\geq 0.\smallskip  \label{WE}
\end{eqnarray}
where $\delta \left( {}\right) $ is Dirac\'{}s delta. Hence the fluctuation
of the energy is 
\begin{equation}
\sqrt{\left\langle U^{2}\right\rangle -\left\langle U\right\rangle ^{2}}%
=\left\langle U\right\rangle =\frac{1}{2}
\rlap{\protect\rule[1.1ex]{.325em}{.1ex}}h%
\omega _{0}.  \label{fluctE}
\end{equation}
The distributions of positions and momenta, eqs.$\left( \ref{Wx}\right) $
and $\left( \ref{Wv}\right) $ agree with the QM predictions, but this is not
the case for the energy because QM predicts a sharp energy, in disagreement
with the SED eq.$\left( \ref{WE}\right) .$ Below we shall study this
discrepancy, that is very relevant for our realistic interpretation of
quantum theory.

Eqs.$\left( \ref{Wx}\right) $ and $\left( \ref{Wv}\right) $ show that the
Heisenberg uncertainty relations, 
\begin{equation}
\Delta x\Delta p\geq 
\rlap{\protect\rule[1.1ex]{.325em}{.1ex}}h%
/2,  \label{Heisineq}
\end{equation}
appear in a natural way in SED. Indeed the probability distributions eqs.$%
\left( \ref{Wx}\right) $ and $\left( \ref{Wv}\right) $ correspond to what in
quantum language is called a ``minimum uncertainty wavepacket'', that is the
quantum state where the Heisenberg inequality, eq.$\left( \ref{Heisineq}%
\right) ,$ saturates i. e. it becomes an equality.

Calculating the corrections due to the finite value of the parameter $\tau $
in eqs.$\left( \ref{2.3}\right) $ to $\left( \ref{energy}\right) $ is
straightforward although lenghty\cite{S74},\cite{dice} and it will not be
reproduced here. A relevant point is that the correction is not analytical
in $\tau $ (or in the fine structure constant $\alpha ),$ but the leading
term agrees with the radiative corrections of quantum electrodynamics (Lamb
shift). An advantage of the SED calculation is that the radiative
corrections (to the nonrelativistic treatment) may be got exactly whilst in
quantum electrodynamics the required perturbative techniques allow only an
expansion in powers of $\tau $ (or $\alpha ),$ once a ultraviolet cut-off is
introduced. In any case the radiative corrections depend on the high
frequency region of integrals like eq.$\left( \ref{oscilspectrum}\right) ,$
where the non-relativistic approximation breaks down. Therefore the
calculation of these corrections has a purely academic interest.

\subsection{Comparison between the stationary state in SED and the ground
state in QM}

A conclusion of the study of the stationary state of the oscillator in SED
is that it is rather similar to the ground state of the oscillator en QM.
Indeed the probability distribution of positions and momenta in the
stationary state of SED agree with the predictions of QM for the ground
state, in the limit $\tau \rightarrow 0$, eqs.$\left( \ref{Wx}\right) $ and $%
\left( \ref{Wv}\right) ,$ whilst the corrections for finite $\tau $, that
depend on the small quantity $\tau \omega _{0},$ correspond to the radiative
corrections of quantum electrodynamics. However the probability distribution
of the energy does not agree with QM. In the following I study more
carefully this discrepancy.

\subsubsection{John von Neumann\'{}s theorem against hidden variables}

Firstly I should mention that the conflict between the QM prediction and the
SED eq.$\left( \ref{WE}\right) $ is an example of the general argument used
by von Neumann\cite{von Neumann4} in his celebrated theorem of 1932 proving
that hidden variable theories are incompatible with QM. That theorem
prevented research in hidden variables theories until Bell\'{}s rebuttal in
1966\cite{BellRMP4}. J. von Neumann starts with the assumption that any
linear relation between quantum observables should correspond to a similar
linear relation between the possible (dispersion free) values in a
hypothetical hidden variables theory. In our case the energy $U$ is a linear
combination of $v^{2}$ and $x^{2}.$ Thus as the energy predicted by quantum
mechanics, $U=
\rlap{\protect\rule[1.1ex]{.325em}{.1ex}}h%
\omega _{0}/2,$ is sharp, any pair of values of $v^{2}$ and $x^{2}$ in the
hidden variables theory should fulfil, according to von Neumann\'{}s
hypothesis, 
\begin{equation}
m(v^{2}+\omega _{0}^{2}x^{2})=
\rlap{\protect\rule[1.1ex]{.325em}{.1ex}}h%
\omega _{0},  \label{linear}
\end{equation}
which is not compatible with the distributions eqs.$\left( \ref{Wx}\right) $
and $\left( \ref{Wv}\right) $ (for instance the possible value $v^{2}=2 
\rlap{\protect\rule[1.1ex]{.325em}{.1ex}}h%
\omega _{0}/m$ is incompatible with eq.$\left( \ref{linear}\right) $ because
it would imply $x^{2}\geq 0).$ Bell's rebutted von Neumann pointing out that
the contradiction only arises when two of the quantum obervables involved do
not commute and in this case the measurement of the three observables should
be made in, at least, two different experiments. Thus a contextual hidden
variables theory is possible, that is a theory where it is assumed that the
value obtained in the measurement depends on both the state of the observed
system and the full experimental context.

\subsubsection{The apparent contradiction between QM and SED}

In our case the apparent contradiction between SED eq.$\left( \ref{WE}%
\right) $ and the QM prediction of a sharp energy dissapears if we take into
account how\emph{\ }the energy of a state is defined \emph{operationally }%
(i. e. how it may be measured.) In SED the stationary state corresponds to a
dynamical equilibrium between the oscillator and the ZPF. Checking
empirically whether a dynamical equilibrium exists requires a long time,
ideally infinite time. If we define the energy of the oscillator in
equilibrium as the average over an infinite time, it would be obviously
sharp. In fact the probability distribution of the ``mean energies over time
intervals of size $\Delta t$ $"$ has a smaller dispersion as greater is $%
\Delta t,$ and will be dispersion free in the limit $\Delta t\rightarrow
\infty .$ Thus it is natural to assume that the ground state energy as
defined by QM actually corresponds to measurements made over infinitely long
times. This fits fairly well with the quantum energy-time uncertainty
relation 
\begin{equation}
\Delta U\Delta t\geq 
\rlap{\protect\rule[1.1ex]{.325em}{.1ex}}h%
/2,  \label{energytime}
\end{equation}
which predicts that the measured energy does possess a dispersion $\Delta U$
if the measurement involves a finite time $\Delta t$. Thus no contradiction
exists between SED and QM for the energy in the ground state.

It is remarkable that QM and SED lead to the same result via rather
different paths. In fact in QM the state vector of the ground state of a
system is an eigenstate of the Hamiltionian, which implies a nil dispersion
of the state energy, but the uncertainty relation gives rise to some
uncertainty for any actual measurement. This leads us to propose that \emph{%
the ground state of a physical system in QM corresponds to a dynamical
equilibrium between emission of radidation to the vacuum fields and
absorption from them.} The instantaneous energy is a badly defined concept.
Indeed the SED distribution eq.$\left( \ref{WE}\right) $ derives from the
(classical) definition of total energy in terms of positions and momenta,
but it does not possesses any operational (measurable) meaning.

\subsection{Spectrum of the light emitted or absorbed by the SED oscillator}

There is another trivial agreement between the SED and QM predictions for
the oscillator, namely the spectrum of emitted or absorbed light. In fact
the standard quantum method to derive the spectrum of a system starts
solving the stationary Schr\"{o}dinger equation and then calculating the
frequencies using the rule 
\[
\omega _{jk}=\frac{E_{j}-E_{k}}{
\rlap{\protect\rule[1.1ex]{.325em}{.1ex}}h%
}=(j-k)\omega _{0}, 
\]
where the eigenvalues of the oscillator Hamiltonian, $E_{n}=n 
\rlap{\protect\rule[1.1ex]{.325em}{.1ex}}h%
\omega _{0},$ have been taken into account. However in the oscillator there
is a selection rule that, within the electric dipole approximation, forbids
transitions except if $j-k=\pm 1,$ whence the spectrum has a single
frequency that agrees with the classical one. Actually the spectrum contains
also the frequencies $n\omega _{0}$, that correspond to electric multipole
transitions, although these transitions have low probability. The multipoles
of the fundamental frequency may be found also in SED calculations if the
electric dipole approximation is not made, that is if the following Lorentz
force is substituted for the last term of eq.$\left( \ref{ode}\right) $%
\[
F_{x}=e\left[ \mathbf{E+}\frac{\mathbf{\dot{r}}}{c}\times \mathbf{B}\right]
_{x}, 
\]
$\mathbf{E}$ and $\mathbf{B}$ being the electric and magnetic fields of the
ZPF. To be consistent the other terms of eq.$\left( \ref{ode}\right) $
should be also changed to become relativistic in order to be consistent.
Then the differential equation of motion becomes nonlinear and it is far
more difficult to solve, but this may be achieved numerically and good
agreement with quantum predictions is obtained for the spectrum of emitted
or absorbed light\cite{Huang}.

SED may also offer intuitive pictures for cavity quantum electrodynamics, a
well established experimental field of research\cite{dice}. An atom in a
cavity get modified its properties, in particular its lifetime. In fact the
atom does not decay if the modes having the frequency of the emitted
radiation are not possible inside the cavity. In the quantum treatment the
intriguing question is how the atom ``knows'' in advance that it should not
decay in these conditions. In SED the explanation is trivial: spontaneous
decay is actually stimulated by appropriate modes of the ZPF, and the modes
required for the stimulation do not exist inside the cavity. For instance in
an early experiment by Haroche et al.\ref{Haroche} the excited atoms
propagates between two metallic mirrors separated by 1.1 $\mu m$\ for about
13 natural lifetimes without appreciable decay. The experiment involved a
small applied magnetic field in order to demonstrate the anisotropy of
spontaneous emission between mirrors. This experiment has been studied
within SED via modelling the atom by a harmonic oscillator whence the
empirical results have been reproduced quantitatively, but I will not review
that work here\cite{Humba2}, \cite{Humba}.

\subsection{Lessons for a realistic interpretation of quantum theory}

\textit{Our study of a particle in a potential well shows that inclusion of
the vacuum random electromagnetic field leads to predictions resembling
those of quantum electrodynamics. }

\textit{The quantum ground state of a particle in a potential well
corresponds to a stationary state of a particle performing a highly
irregular (stochastic) motion driven by vacuum fields.}

\textit{The spectrum of the field (or the energy per unit volume and unit
frequency interval) determines the spectrum for the motion of the particle.
It is such that the quadratic mean coordinate position and the quadratic
mean momentum agree with quantum predictions and fulfil the Heisenberg
uncertainty relations. These are here interpreted as a consequence of the
(unavoidable) random motion of the particles. }

\textit{No contradiction arises between the exponential distribution of
energy in SED and the sharp energy in QM. They are different operational
definitions. The former refers to the instantaneous energy }(\textit{or the
mean in a small time interval) but the latter to the mean over an infinite
(or very large) time interval. This difference is a good illustration for
the flaw in the celebrated von Neumann theorem against hidden variables in
quantum mechanics.}

\textit{Radiative corrections (e. g. Lamb shift) appear naturally in SED
with a transparent interpretation, i. e. as a consequence of the interaction
between the charged particle and the real vacuum fields.}

\section{Time-dependent properties of oscillators and free particles}

\subsection{Evolution of the dynamical variables}

In Newtonian mechanics the study of the evolution consists of finding the
position as a function of time for given initial conditions, that is initial
positions and velocities of the particles involved. Thus the evolution
describes a curve in phase space parametrized by time. If there are forces
not fully known, which we represent as noise, all we may get is the
evolution of the probability distribution in phase space with given initial
conditions, i. e. either a point or a probability distribution in phase
space. This is the case for the oscillator in SED that we study in the
following. In order to calculate the evolution of the oscillator it is
convenient to start anew from the equation of motion, eq.$\left( \ref{ode}%
\right) $. We shall work to lowest nontrivial order in the small parameter $%
\tau $ (see eq.$\left( \ref{gamma}\right) )$ Thus we may approximate the
third order eq.$\left( \ref{ode}\right) $ by another one of second order
substituting $-m\tau \omega _{0}^{2}\dot{x}$ for $m\tau \stackrel{...}{x}$
on its right side. That is writing 
\begin{equation}
m\stackrel{..}{x}=-m\omega _{0}^{2}x-m\tau \omega _{0}^{2}\dot{x}+eE\left(
t\right) ,  \label{ode1}
\end{equation}
which agrees with eq.$\left( \ref{ode}\right) $ to first order in $\tau .$
This second order equation in $x(t)$ is equivalent to two coupled stochastic
differential equations of Langevin type, in the variables $x(t)$ and $\dot{x}%
(t).$ We used eq.$\left( \ref{ode}\right) ,$ rather than $\left( \ref{ode1}%
\right) $ because the former was more appropriate for the study of radiative
corrections than the latter.

A convenient vay to study the motion of the oscillator in SED consists of
introducing new variables, $a(t)$\ and $b(t),$ as follows 
\begin{eqnarray}
x\left( t\right) &=&a(t)\cos \left( \omega _{0}t\right) +b(t)\sin \left(
\omega _{0}t\right) +\xi (t),  \label{xab} \\
\dot{x}\left( t\right) &=&-a(t)\omega _{0}\sin \left( \omega _{0}t\right)
+b(t)\omega _{0}\cos \left( \omega _{0}t\right) +\dot{\xi}(t).  \nonumber
\end{eqnarray}
The rapidly fluctuating quantity $\xi (t)$\ is related to the high frequency
part of the spectrum $S_{x}\left( \omega \right) $\ (see comment after eq.$%
\left( \ref{2.4}\right) $) and it will be ignored in the following. The
variables $a$ and $b$ are constants of the motion in the classical
mechanical oscillator and they are slowly varying functions of time, with
typical variation time $1/(\tau \omega _{0}^{2})>>\omega _{0}^{-1},$ see
below. An alternative to eq.$\left( \ref{xab}\right) $ would be to write the
coordinate in terms of the amplitude, $c(t),$ and the phase, $\phi \left(
t\right) ,$ both slowly varying with time, that is 
\[
x\left( t\right) =c(t)\cos \left[ \omega _{0}t+\phi \left( t\right) \right]
+\xi (t), 
\]
but the choise eq.$\left( \ref{xab}\right) $ is more easy to solve. At the
initial time, $t=0$, the parameters $a(t)$\ and $b(t)$ are easily related to
the initial position, $x_{0},$ and momentum, $p_{0}$, that is 
\begin{equation}
a\left( t_{0}\right) =x_{0},b\left( t_{0}\right) =\frac{\dot{x}_{0}}{\omega
_{0}}=\frac{p_{0}}{m\omega _{0}}.  \label{xpab}
\end{equation}

Calculating the evolution of the variables $a(t)$\ and $b(t)$ simplifies if
we introduce a complex function $z\left( t\right) $ such that 
\begin{eqnarray}
x(t) &=&\func{Re}\left[ z(t)\exp (-i\omega _{0}t)\right] ,  \label{z} \\
\func{Re}z(t) &=&a(t),\func{Im}z(t)=b(t).  \nonumber
\end{eqnarray}
The function $z(t)$ is slowly varying and therefore we may neglect its
second (first) derivative in the term of order $0$ (order $\tau $) in the
equation that results from inserting eq.$\left( \ref{z}\right) $ into eq.$%
\left( \ref{ode1}\right) .$ This gives 
\begin{equation}
-2im\omega _{0}\dot{z}\left( t\right) =im\tau \omega _{0}^{3}z\left(
t\right) +eE\left( t\right) \exp \left( i\omega _{0}t\right) .  \label{z1}
\end{equation}
The solution of this equation is trivial and we get 
\[
z\left( t\right) =\exp \left( -\frac{1}{2}\tau \omega _{0}^{3}t\right)
\left[ z(0)+\frac{ie}{2m\omega _{0}}\int_{0}^{t}E(t^{\prime })\exp \left( 
\frac{1}{2}\tau \omega _{0}^{2}t^{\prime }+i\omega _{0}t^{\prime }\right)
dt^{\prime }\right] . 
\]
Hence it is easy to get the ensemble average, or expectation, of $z\left(
t\right) $ taking into account that $E(t)$ is a stochastic process with zero
mean. We get 
\begin{equation}
\left\langle z\left( t\right) \right\rangle =\exp \left( -\frac{1}{2}\tau
\omega _{0}^{2}t\right) \left\langle z\left( 0\right) \right\rangle .
\label{z0}
\end{equation}
Also we may obtain the following quadratic mean 
\begin{equation}
\left\langle \left| z\left( t\right) -\exp \left( -\frac{1}{2}\tau \omega
_{0}^{3}t\right) z\left( 0\right) \right| ^{2}\right\rangle =\frac{e^{2}}{%
4m^{2}\omega _{0}^{2}}\exp \left( -\tau \omega _{0}^{3}t\right) F,
\label{z2}
\end{equation}
where 
\begin{equation}
F\equiv \int_{0}^{t}dt^{\prime }\int_{0}^{t}dt^{\prime \prime }\left\langle
E(t^{\prime })E(t^{\prime \prime })\right\rangle \exp \left[ \left( \frac{1}{%
2}\tau \omega _{0}^{2}+i\omega _{0}\right) t^{\prime }+\left( \frac{1}{2}%
\tau \omega _{0}^{2}-i\omega _{0}\right) t^{\prime \prime }\right] .
\label{F}
\end{equation}
The selfcorrelation of the process $E(t)$ is the Fourier transform of the
spectral density, that is 
\begin{equation}
\left\langle E(t^{\prime })E(t^{\prime \prime })\right\rangle =\frac{1}{2}%
\int_{-\infty }^{\infty }\frac{2}{3\pi c^{3}}
\rlap{\protect\rule[1.1ex]{.325em}{.1ex}}h%
\omega ^{3}\exp \left[ i\omega \left( t^{\prime \prime }-t^{\prime }\right)
\right] d\omega ,  \label{EE}
\end{equation}
where eq.$\left( \ref{Espectrum}\right) $ has been taken into account. I
stress that the process $E(t)$ is stationary and assumed ergodic, whence
time average (over an infinite time) and ensemble average agree. However
here we do not assume that $x(t)$ is stationary but we are investigating its
time dependence. If eq.$\left( \ref{EE}\right) $ is put in eq.$\left( \ref{F}%
\right) $ and the integrals in $t^{\prime }$ and $t^{\prime \prime }$
performed we get 
\begin{eqnarray*}
F &=&\frac{
\rlap{\protect\rule[1.1ex]{.325em}{.1ex}}h%
}{3\pi c^{3}}\int_{-\infty }^{\infty }\omega ^{3}d\omega \frac{4}{\tau
^{2}\omega _{0}^{4}+4\left( \omega -\omega _{0}\right) ^{2}}\left| \exp
\left( \frac{1}{2}\tau \omega _{0}^{2}+i\omega _{0}-i\omega \right)
t-1\right| ^{2} \\
&=&\frac{
\rlap{\protect\rule[1.1ex]{.325em}{.1ex}}h%
}{3\pi c^{3}}\int_{-\infty }^{\infty }\frac{4\omega ^{3}}{\tau ^{2}\omega
_{0}^{4}+4\left( \omega -\omega _{0}\right) ^{2}}\times \\
&&\times \left[ \exp (\tau \omega _{0}^{2}t)+1-2\exp (\frac{1}{2}\tau \omega
_{0}^{2}t)\cos \left( \omega t-\omega _{0}t\right) \right] d\omega .
\end{eqnarray*}
The integral in $\omega $ is ultraviolet divergent but the contribution of
the high frequencies will be ignored here (see comment after eq.$\left( \ref
{xab}\right) ).$ Thus taking into account that $\tau \omega _{0}<<1,$ the
overwhelming contribution to the integral comes from frequencies $\omega
\simeq \omega _{0}$ and the integral may be approximated putting $\omega
^{3}=\omega _{0}^{3}$ whence we obtain 
\begin{equation}
\left\langle \left| z\left( t\right) -\exp \left( -\frac{1}{2}\tau \omega
_{0}^{3}t\right) z\left( 0\right) \right| ^{2}\right\rangle \simeq \frac{%
\rlap{\protect\rule[1.1ex]{.325em}{.1ex}}h%
}{2m\omega _{0}^{2}}\left[ 1-\exp \left( -\tau \omega _{0}^{2}t\right)
\right] .  \label{zz}
\end{equation}
Taking eq.$\left( \ref{z}\right) $ into account we may separate the real and
the imaginary parts in eqs.$\left( \ref{z0}\right) $ and $\left( \ref{zz}%
\right) .$ Thus we obtain information about the evolution of the classical
constants of the motion $a=Re$ $z$ and $b=Im$ $z$, that is 
\begin{eqnarray}
\left\langle a\left( t\right) \right\rangle &=&\exp \left( -\frac{1}{2}\tau
\omega _{0}^{2}t\right) \left\langle a\left( 0\right) \right\rangle ,
\label{a12} \\
\left\langle \left[ a(t)-\left\langle a\left( t\right) \right\rangle \right]
^{2}\right\rangle &=&\left\langle \left[ a(t)-\exp \left( -\frac{1}{2}\tau
\omega _{0}^{3}t\right) a(0)\right] ^{2}\right\rangle  \nonumber \\
&\simeq &\frac{
\rlap{\protect\rule[1.1ex]{.325em}{.1ex}}h%
}{4m\omega _{0}^{2}}\left[ 1-\exp \left( -\tau \omega _{0}^{2}t\right)
\right] ,  \nonumber
\end{eqnarray}
and similar for $b(t)$ because both variables, $a(t)$\ and $b(t),$ have
similar contributions to eq.$\left( \ref{zz}\right) ,$ which follows from
their roles in eqs.$\left( \ref{xab}\right) .$ Eqs.$\left( \ref{a12}\right) $
and the similar ones for $b(t)$ mean that for any initial distribution of
positions and momenta, the oscillator will arrive at a distribution
corresponding to the stationary state (corresponding to the quantum ground
state).

\subsection{Diffusion of the probability density in phase space}

It is possible to derive differential equations for the probability
densities of $a$ and $b.$ They have the form of Fokker-Planck (or diffusion
) equation like 
\[
\frac{\partial \rho (a,t)}{\partial t}=\frac{\partial }{\partial a}(A\rho )+%
\frac{\partial ^{2}}{\partial a^{2}}(D\rho ). 
\]
The coefficients of drift, $A,$ and diffusion, $D,$ may be calculated from
eqs.$\left( \ref{a12}\right) $ as follows 
\[
A=\lim_{t\rightarrow \infty }\frac{\left\langle a(t)\right\rangle }{t}=-%
\frac{1}{2}\tau \omega _{0}^{2},D=\lim_{t\rightarrow \infty }\frac{%
\left\langle \left[ a(t)-\left\langle a\left( t\right) \right\rangle \right]
^{2}\right\rangle }{t}=\frac{
\rlap{\protect\rule[1.1ex]{.325em}{.1ex}}h%
\tau }{4m}. 
\]
whence the Fokker-Planck equation reads 
\begin{equation}
\frac{\partial \rho (a,t)}{\partial t}=\frac{1}{2}\tau \omega _{0}^{2}\frac{%
\partial }{\partial a}(a\rho )+\frac{
\rlap{\protect\rule[1.1ex]{.325em}{.1ex}}h%
\tau }{4m}\frac{\partial ^{2}\rho }{\partial a^{2}},  \label{FP}
\end{equation}
and a similar one with $b$ substituted for $a$. I must point out that the
stochastic processes $a(t)$ and $b(t)$ are correlated. In fact their
crosscorrelation may be easily obtained as follows 
\begin{eqnarray*}
\left\langle a(t)b(t^{\prime })\right\rangle &=&\left\langle \left[ x\left(
t\right) \cos \left( \omega _{0}t\right) -\frac{p(t)}{m\omega _{0}}\sin
\left( \omega _{0}t\right) \right] \left[ x(t^{\prime })\cos \left( \omega
_{0}t^{\prime }\right) +\frac{p(t^{\prime })}{m\omega _{0}}\sin \left(
\omega _{0}t^{\prime }\right) \right] \right\rangle \\
&=&-\frac{1}{2}
\rlap{\protect\rule[1.1ex]{.325em}{.1ex}}h%
\sin \left[ \omega _{0}\left( t^{\prime }-t\right) \right] \exp (-\tau
\omega _{0}^{2}\left| t^{\prime }-t\right| )=-\left\langle b(t)a(t^{\prime
})\right\rangle ,
\end{eqnarray*}
where we have taken into account eqs.$\left( \ref{xab}\right) .$ Due to this
correlation a joint probability density $\rho (a,$ $b,t)$ cannot be got as a
product of the individual densities. Consequently deriving the differential
equation for $\rho (a,$ $b,t)$ is involved and it will be not made here.

The conclusion of our calculation is that the classical constants of the
motion, like the parameters $a$ and $b$ or the energy, $U$, perform a slow
random motion with typical relaxation time $1/\left( \tau \omega
_{0}^{2}\right) .$ In particular the energy is related to these parameters
as follows\textrm{\ } 
\begin{equation}
U(t)=\frac{1}{2}m\omega _{0}^{2}x(t)^{2}+\frac{1}{2}m\dot{x}(t)^{2}=\frac{1}{%
2}m\omega _{0}^{2}\left[ a(t)^{2}+b(t)^{2}\right] ,  \label{Uab}
\end{equation}
where eqs.$\left( \ref{xab}\right) $ have been taken into account. The
change of the classical constants of the motion of the oscillator in SED is
obviously due to the two latter terms of eq.$\left( \ref{ode}\right) .$\ The
term involving the vacuum field produces diffusion, characterized by $D$,
and the radiation reaction term gives rise to drift, characterized by $A$.
The diffusion rate is independent of the particle\'{}s velocity whence $D$
is a constant, but the drift increases with the velocity with the result
that $A$ is proportional to $a.$ The effect of the diffusion is reduced to
some extent by the drift, with the consequence that the probability
densities remain localized. In fact, when time increases indefinitely the
densities approach the stationary solution studied in section 2. In
particular the stationary solution of eq.$\left( \ref{FP}\right) $ (with
normalized $\rho ,$\ which implies that $\rho $\ vanishes for $a\rightarrow
\pm \infty )$\ is 
\[
\rho =\sqrt{\frac{
\rlap{\protect\rule[1.1ex]{.325em}{.1ex}}h%
}{\pi m\omega _{0}^{2}}}\exp \left( -\frac{
\rlap{\protect\rule[1.1ex]{.325em}{.1ex}}h%
}{m\omega _{0}^{2}}a^{2}\right) , 
\]
and similar for $b$. The energy tends to the density given by eq.$\left( \ref
{WE}\right) .$

\subsection{States of the oscillator in stochastic electrodynamics and in
quantum mechanics}

Every nonnegative definite function in phase space may be taken as an
initial probability density and thus be considered a state of the SED
oscillator. We see that the set of states in SED is quite different from the
set of states in QM (given by a density operator each). In particular the
pure states in SED are those whose initial conditions correspond to points
in phase space, whilst the pure states in QM correspond to state vectors (or
wave functions). The comparison between SED and QED becomes more clear if we
define the quantum states by means of functions in phase space, which might
be achieved via the Wigner function formalism, to be revisited in Chapter 5.
Thus in the case of pure quantum states it is obvious that only a small
fraction of them correspond to states in SED. Actually, asides from the
ground state there are only two interesting pure quantum states that
correspond precisely to SED states, namely coherent states and squeezed
states. The latter are relevant in case of radiation (squeezed states of
light), but not so much for matter oscillators and they will not be studied
here.

Coherent states in SED appear as solutions of the oscillator eq.$\left( \ref
{ode}\right) $ obtained by combining the stationary solution of the equation
with the general solution of the homogenous equation, that is 
\[
\stackrel{..}{x}+\omega _{0}^{2}x+\tau \omega _{0}^{2}\stackrel{.}{x}%
=0\Rightarrow x\simeq A\cos \left( \omega _{0}t+\phi \right) \exp \left(
-\tau \omega _{0}t\right) , 
\]
where I have approximated $\stackrel{...}{x}\simeq -\omega _{0}^{2}\stackrel{%
.}{x}$ and neglected a small shift, of order $\tau ,$ in the frequency $%
\omega _{0}.$ Hence, taking eq.$\left( \ref{Wx}\right) $ into account, we
see that the solution of eq.$\left( \ref{ode}\right) $ leads to the
following time dependent probability distribution of positions

\begin{equation}
W\left( x,t\right) \simeq \sqrt{\frac{m\omega _{0}}{\pi 
\rlap{\protect\rule[1.1ex]{.325em}{.1ex}}h%
}}\exp \left[ -\frac{m\omega _{0}}{2
\rlap{\protect\rule[1.1ex]{.325em}{.1ex}}h%
}\left[ x-A\cos \left( \omega _{0}t+\phi \right) \exp \left( -\tau \omega
_{0}^{2}t\right) \right] ^{2}\right] ,  \label{general}
\end{equation}
which contains two integration constants, $A$ and $\phi .$ It must be
stressed that this expression for the probability density derives from eq.$%
\left( \ref{ode}\right) $ and the ZPF spectrum eq.$\left( \ref{Espectrum}%
\right) $ with the approximation of putting $\tau \rightarrow 0$ except in
the exponential decay. It may be seen that when $\tau =0$ the evolution of
the position probability density eq.$\left( \ref{general}\right) $ fully
agrees with the one of the coherent states of quantum mechanics, whilst the
expression for finite $\tau $ contains the most relevant contribution of the
radiative corrections of quantum electrodynamics to these states (a decay
towards the stationary state with relaxation time $\left( \tau \omega
_{0}^{2}\right) ^{-1}$)\cite{S74}.

\textit{In summary we see that a few states of the oscillator in SED
correspond to pure quantum states in a phase-space representation. But no
pure state of SED corresponds to a state of QM. Also most of the pure states
of QM do not correspond to states of SED. However for mixed states the
agreement is greater, and actually all (mixed) quantum states possessing a
positive Wigner function closely correspond to mixed states of SED with the
same phase-space distribution. }

\textit{In spite of these differences the quantum theory of the harmonic
oscillator admits a realistic interpretation via SED, provided that the same
predictions may be obtained for actual experiments. In particular the
quantum states, solutions of Schr\"{o}dinger equation, might be just
mathematical auxiliary functions used in the QM formalism but not required
in the SED approach for the prediction of the same results. }

\subsection{The free particle}

For a free particle the differential equation of motion is like the
oscillator's eq.$\left( \ref{ode}\right) $ with $\omega _{0}=0,$ that is 
\begin{equation}
m\stackrel{..}{x}=m\tau \stackrel{...}{x}+eE\left( t\right) .
\label{odefree}
\end{equation}
This fact may suggest studying the free particle as the limit of an
oscillator whose characteristic frequency decreases to zero. However this
method is not appropriate because there is a qualitative difference between
the two systems. In fact the motion in the oscillator is always bound which
is not the case for the free particle. Thus we shall study the motion of the
free particle starting from eq.$\left( \ref{odefree}\right) .$ It is a third
order equation and therefore has three independent solutions, but one of
them is \textit{runaway }that is the energy increases without limit, which
is physically nonsense. The reason is that the radiation reaction term, the
former term on the right side of eq.$\left( \ref{odefree}\right) $, is a
linearized approximation not valid for a free particle (in the oscillator
the runaway solution is effectively cut-off by the potential and the
approximation of eq.$\left( \ref{ode}\right) $ is good enough). Thus we
shall substitute the following integro-differential equation for eq.$\left( 
\ref{odefree}\right) $%
\[
\stackrel{..}{x}=-\frac{e}{m\tau }\exp \left( \frac{t}{\tau }\right)
\int_{t}^{\infty }E\left( t^{\prime }\right) \exp \left( -\frac{t^{\prime }}{%
\tau }\right) dt^{\prime }. 
\]
It has the same solutions as eq.$\left( \ref{odefree}\right) $ except the
runaway ones. Hence, it is trivial to get the following equations of
evolution for the velocity and the coordinate, respectively, that is 
\begin{eqnarray}
v(t) &=&v_{0}-\frac{e}{m\tau }\int_{0}^{t}\exp \left( \frac{s}{\tau }\right)
ds\int_{s}^{\infty }E\left( u\right) \exp \left( -\frac{u}{\tau }\right) du,
\label{dispv} \\
x(t) &=&x_{0}+v_{0}t-\frac{e}{m\tau }\int_{0}^{t}ds\int_{0}^{s}\exp \left( 
\frac{u}{\tau }\right) du\int_{u}^{\infty }E\left( w\right) \exp \left( -%
\frac{w}{\tau }\right) dw,  \nonumber
\end{eqnarray}
where $x_{0}$ is the initial position and $v_{0}$ the initial velocity at
time $t=0$. Hence, taking into account that the ensemble average of $E\left(
t\right) $ is zero, it is trivial to get the mean position and velocity of a
particle, that is 
\begin{equation}
\left\langle x\left( t\right) \right\rangle =x_{0}+v_{0}t.  \label{xvt}
\end{equation}

The most interesting quantities are the dispersions of velocity and position
with time. The velocity dispersion may be got from the first eq.$\left( \ref
{dispv}\right) $ putting $v_{0}=0.$ We obtain, taking eq.$\left( \ref{gamma}%
\right) $ into account, 
\begin{eqnarray}
\left\langle v\left( t\right) ^{2}\right\rangle &=&\frac{3c^{3}}{2m\tau }%
\int_{0}^{t}\exp \left( \frac{s}{\tau }\right) ds\int_{s}^{\infty }\exp
\left( -\frac{u}{\tau }\right) du  \label{vv} \\
&&\times \int_{0}^{t}\exp \left( \frac{s^{\prime }}{\tau }\right) ds^{\prime
}\int_{s}^{\infty }\exp \left( -\frac{u^{\prime }}{\tau }\right) du^{\prime
}\left\langle E\left( u\right) E\left( u^{\prime }\right) \right\rangle . 
\nonumber
\end{eqnarray}
The $E(t)$ selfcorrelation is the Fourier transform of the spectrum (see eq.$%
\left( \ref{deltaxx}\right) ),$ that is 
\begin{eqnarray}
\left\langle E\left( u\right) E\left( u^{\prime }\right) \right\rangle
&=&\int_{0}^{\infty }S_{x}\left( \omega \right) \cos \left[ \omega \left(
u-u^{\prime }\right) \right] d\omega  \nonumber \\
&=&\frac{1}{2}\int_{-\infty }^{\infty }\left| S_{x}\left( \omega \right)
\right| \exp \left[ i\omega \left( u-u^{\prime }\right) \right] d\omega .
\label{Ess}
\end{eqnarray}
I point out that this relation is correct because $E(t)$ is a stationary
process, but it is not possible to get eq.$\left( \ref{vv}\right) $ from the
spectrum of $v\left( t\right) $ because in the free particle case we cannot
get the spectrum of $v(t)$ from that of $E(t)$ (as we made in the derivation
of eq.$\left( \ref{oscilspectrum}\right) $ for the equilibrium state of
oscillator, where both $x(t)$ and $v(t)$ are stationary processes).
Inserting eq.$\left( \ref{Ess}\right) $ in eq.$\left( \ref{vv}\right) $ we
get, after changing the order of the integrations, 
\begin{eqnarray*}
\Delta v^{2} &\equiv &\left\langle v\left( t\right) ^{2}\right\rangle =\frac{%
e^{2}}{2m^{2}\tau ^{2}}\int_{-\infty }^{\infty }\left| S_{x}\left( \omega
\right) \right| d\omega \left| \int_{0}^{t}\exp \left( \frac{s}{\tau }%
\right) ds\int_{s}^{\infty }\exp \left( -\frac{u}{\tau }+i\omega u\right)
du\right| ^{2} \\
&=&\frac{
\rlap{\protect\rule[1.1ex]{.325em}{.1ex}}h%
}{2\pi m\tau }\int_{-\infty }^{\infty }\frac{\left| \omega \right| d\omega }{%
\omega ^{2}+\tau ^{-2}}\left| 1-\exp (i\omega t)\right| ^{2}=\frac{%
\rlap{\protect\rule[1.1ex]{.325em}{.1ex}}h%
\tau }{\pi m}\int_{0}^{\infty }\frac{\omega d\omega }{1+\tau ^{2}\omega ^{2}}%
\left| 1-\cos (\omega t)\right| .
\end{eqnarray*}
Thus the velocity dispersion gives an ultraviolet divergent integral that
may be made convergent by introducing a cut-off frequency $\omega _{c}$.
Thus we get 
\begin{eqnarray}
\Delta v^{2} &=&\frac{
\rlap{\protect\rule[1.1ex]{.325em}{.1ex}}h%
\tau }{\pi m}\int_{0}^{\omega _{c}}\frac{\omega d\omega }{1+\tau ^{2}\omega
^{2}}\left[ 1-\cos (\omega t)\right]  \label{4.10} \\
&\sim &\frac{
\rlap{\protect\rule[1.1ex]{.325em}{.1ex}}h%
}{2\pi m\tau }\left[ \log (1+\omega _{c}^{2}\tau ^{2})+\frac{\tau ^{2}}{t^{2}%
}\right] ,\text{ for }t>>\tau .  \nonumber
\end{eqnarray}
The dispersion $\Delta v$ becomes rapidly independent of $t$, but greater
than the velocity of light. In order that $\Delta v<c$ we must have 
\[
\omega _{c}<\sqrt{\frac{3\pi }{\alpha }}\frac{mc^{2}}{%
\rlap{\protect\rule[1.1ex]{.325em}{.1ex}}h%
}\approx 0.2\frac{c}{\lambda _{C}}<<\frac{1}{\tau }, 
\]
$\lambda _{C}$ being the Compton wavelength. This implies that eq.$\left( 
\ref{4.10}\right) $ may be rewritten 
\begin{equation}
\Delta v^{2}\sim \frac{
\rlap{\protect\rule[1.1ex]{.325em}{.1ex}}h%
\tau \omega _{c}^{2}}{2\pi m}+\frac{
\rlap{\protect\rule[1.1ex]{.325em}{.1ex}}h%
}{2\pi m\tau t^{2}},\text{ for }t>>\tau .  \label{4.10a}
\end{equation}
A correct calculation would require a relativistic theory, which will not be
attempted here. Nevertheless the result obtained shows that the particle
performs a random motion with relativistic speed although the mean velocity
remains a constant (see eq.$\left( \ref{xvt}\right) ).$ Also the result
suggests that in a relativistic calculation the most relevant wavelengths
would be those not too far from the Compton one. The increase of the
velocity of a free charged particle by the action of the ZPF has been
proposed as possible origen of the observed ultrahigh-energy X rays coming
to Earth from outside the Solar System\cite{Rueda}.

It is interesting to compare the velocity dispersion of the free particle in
SED with the particle immersed in Rayleigh-Jeans (classical) radiation.
Taking into account that the ZPF and the Rayleigh-Jeans radiation correspond
to $\frac{1}{2}
\rlap{\protect\rule[1.1ex]{.325em}{.1ex}}h%
\omega $ and $kT$ per normal mode, respectively, the replacement $\frac{1}{2}
\rlap{\protect\rule[1.1ex]{.325em}{.1ex}}h%
\omega $ $\rightarrow $ $kT$ in eq.$\left( \ref{4.10}\right) $ leads to 
\begin{eqnarray*}
\Delta v^{2} &=&\frac{\tau kT}{\pi m}\int_{0}^{\omega _{c}}\frac{d\omega }{%
1+\tau ^{2}\omega ^{2}}\left[ 1-\cos (\omega t)\right] \\
&\sim &\frac{kT}{m}\text{ for }t>>\tau .
\end{eqnarray*}
We see that the velocity dispersion of the charged free particle does not
increase indefinitely but becomes, after a long enough time, a constant
corresponding to the kinetic energy $kT/2$ (which is the equipartition of
the energy of classical statistical mechanics.)

The dispersion of position oa the free particle according to SED may be
obtained by a similar method, that is inserting the latter eq.$\left( \ref
{Ess}\right) $ in eq.$\left( \ref{dispv}\right) .$ We obtain 
\begin{eqnarray}
\Delta x^{2} &=&\frac{
\rlap{\protect\rule[1.1ex]{.325em}{.1ex}}h%
\tau }{2\pi m}\int_{-\infty }^{\infty }\frac{\left| \omega \right| d\omega }{%
1+\tau ^{2}\omega ^{2}}\left| \int_{0}^{t}\left[ 1-\exp (i\omega s)\right]
ds\right| ^{2}  \nonumber \\
&=&\frac{
\rlap{\protect\rule[1.1ex]{.325em}{.1ex}}h%
\tau }{\pi m}\int_{0}^{\infty }\frac{\omega d\omega }{1+\tau ^{2}\omega ^{2}}%
\left[ t^{2}-\frac{2t\sin \left( \omega t\right) }{\omega }+\frac{2-2\cos
(\omega t)}{\omega ^{2}}\right]  \nonumber \\
&\simeq &\frac{
\rlap{\protect\rule[1.1ex]{.325em}{.1ex}}h%
\tau \omega _{c}^{2}}{2\pi m}t^{2}+\frac{2
\rlap{\protect\rule[1.1ex]{.325em}{.1ex}}h%
\tau }{\pi m}\left[ \log \left( \frac{t}{\tau }\right) -C-1\right] ,\text{ }%
t>>\tau ,  \label{4.10b}
\end{eqnarray}
where C=0.577... is the Euler constant.

The former term, which dominates at long times, is a consequence of the
velocity dispersion, as may be easily seen by a comparison with eq.$\left( 
\ref{4.10}\right) .$ The (canonical) momentum has no dispersion as shown by
the former eq.$\left( \ref{canonmom}\right) $ when we put $\omega _{0}=0.$
This agrees with the quantum prediction that the momentum of a free particle
is a constant. For a particle with zero canonical momentum, the typical
distance from the original position increases with about one tenth the
velocity of light so that a relativistic treatment would give a quite
different picture.

\textit{The picture that emerges, and gives hints for the interpretation of
the free particle in QM, is as follows. The free particle possesses a
conserved canonical momentum with an associated inertial motion but,
superimposed to this, it has a random motion with a velocity close to that
of light. This produces an apparently contradictory behaviour that derives
from the spectrum,} $S_{E}\left( \omega \right) \varpropto \omega ^{3},$%
\textit{\ of the zeropoint field: At short times the motion is governed by
the high frequencies where} $S_{E}\left( \omega \right) $ \textit{is large
thus inducing a rapid erratic motion, at long times it is governed by the
low frequencies where} $S_{E}\left( \omega \right) $ \textit{is small whence
the memory of the initial velocity is lost very slowly. This fact contrast
with what happens in Brownian motion and what our intuition may suggest,
namely that memory of the initial conditions should be quickly lost. The
special behaviour of diffusion in SED is a consequence of the fact that the
spectrum }$S_{E}\left( \omega \right) $ \textit{is very different from the
popular (Brownian) white noise}, $S_{white}\left( \omega \right) $\textit{\ }%
$\simeq constant.$

\subsection{Commutation rules}

Stochastic electrodynamics also provides a clue for the interpretation of
commutation rules, that is the essential ingredient in the canonical
formulation of quantum mechanics. In fact I will define the commutator at
two times, $t,t^{\prime }$, of a stationary stochastic process, $x\left(
t\right) ,$ via the \textit{sinus} Fourier transform of the spectrum. Then I
shall show that this stochastic commutator applied to the SED oscillator
closely resembles the quantum commutator in the Heisenberg picture of QM.

The introduction of the stochastic commutator is suggested by the Fourier
transform of the spectrum, $S_{x}\left( \omega \right) .$ It consists of two
terms, that is 
\begin{eqnarray}
\int_{0}^{\infty }S_{x}\left( \omega \right) \exp \left[ i\omega \left(
t^{\prime }-t\right) \right] d\omega &=&\left\langle x\left( t\right)
x\left( t^{\prime }\right) \right\rangle +\frac{1}{2}\left[ x\left( t\right)
,x\left( t^{\prime }\right) \right] ,  \nonumber \\
\left\langle x\left( t\right) x\left( t^{\prime }\right) \right\rangle
&=&\int_{0}^{\infty }S_{x}\left( \omega \right) \cos \left[ \omega \left(
t^{\prime }-t\right) \right] d\omega ,  \nonumber \\
\left[ x\left( t\right) ,x\left( t^{\prime }\right) \right]
&=&2i\int_{0}^{\infty }S_{x}\left( \omega \right) \sin \left[ \omega \left(
t^{\prime }-t\right) \right] d\omega ,  \label{comm}
\end{eqnarray}
where the spectrum is defined to be zero for negative frequencies, that is $%
S_{x}\left( \omega \right) =0$ if $\omega <0.$ The real part, $\left\langle
x\left( t\right) x\left( t^{\prime }\right) \right\rangle ,$ is the
selfcorrelation function of the stochastic process so that it is plausible
that the imaginary part, $\left[ x\left( t\right) ,x\left( t^{\prime
}\right) \right] ,$\ is also relevant and we define it as the commutator.
The factor 2 is chosen in order to be similar to the quantum commutator. I
point out that the cosinus Fourier transform of the spectrum is the
selfcorrelation only for stationary processes that are ergodic, as is proved
by the Wiener-Khinchine theorem.

The relation between spectrum and stochastic commutator is also suggested by
the fact that in QM there is a similar relation between the spectrum and the
two-times commutator of the coordinate operator in the Heisenberg picture.
That relation is fulfilled for the ground state of a particle in any
potential well. For the proof here I consider a one-dimensional (quantum)
problem defining the spectrum, $S_{x}\left( \omega \right) ,$ as follows 
\begin{equation}
S_{x}\left( \omega \right) \equiv \sum_{n}\left| \left\langle \psi
_{0}\left| \hat{x}\left( 0\right) \right| \psi _{n}\right\rangle \right|
^{2}\delta \left( \omega -\omega _{0n}\right) ,  \label{sumrule}
\end{equation}
where $\hat{x}$ is the quantum position operator of the particle. The
coefficients of the Dirac's deltas are proportional to the transition
probabilities in QM from the ground state to all possible excited states.
(Although I stress that in QED the deltas are approximations of highly
peaked functions with a finite width when radiative corrections are takent
into account). The analogy with the latter eq.$\left( \ref{comm}\right) $ is
shown as follows. From the Heisenberg equation of motion 
\[
\hat{x}\left( t\right) =\exp (i\hat{H}t/
\rlap{\protect\rule[1.1ex]{.325em}{.1ex}}h%
)\hat{x}\left( 0\right) \exp (-i\hat{H}t/
\rlap{\protect\rule[1.1ex]{.325em}{.1ex}}h%
). 
\]
we may obtain the expectation value of the commutator in the ground state,

\[
\left\langle \left[ \hat{x}\left( 0\right) ,\hat{x}\left( t\right) \right]
\right\rangle =\left\langle \psi _{0}\left| \hat{x}\left( 0\right) \hat{x}%
\left( t\right) \right| \psi _{0}\right\rangle -\left\langle \psi _{0}\left| 
\hat{x}\left( t\right) \hat{x}\left( 0\right) \right| \psi _{0}\right\rangle
. 
\]
After introducing the resolution of the identity between $\hat{x}\left(
0\right) $ and $\hat{x}\left( t\right) $ and between $\hat{x}\left( t\right) 
$ and $\hat{x}\left( 0\right) $ in terms of eigenvectors of the Hamiltonian $%
\hat{H},$ this gives 
\[
\left\langle \left[ \hat{x}\left( 0\right) ,\hat{x}\left( t\right) \right]
\right\rangle =2i\sum_{n}\left| \left\langle \psi _{0}\left| \hat{x}\left(
0\right) \right| \psi _{n}\right\rangle \right| ^{2}\sin \left( \omega
_{0n}t\right) =2i\int S_{x}\left( \omega \right) \sin \left( \omega t\right)
d\omega , 
\]
where in the latter equality we have taken eq.$\left( \ref{sumrule}\right) $
into account. This equality, similar to the stochastic latter eq.$\left( \ref
{comm}\right) ,$ played an important role in the origin of quantum
mechanics. In fact the derivative with respect to $t$ leads to 
\[
2im\sum_{n}\omega _{0n}\left| \left\langle \psi _{0}\left| \hat{x}\left(
0\right) \right| \psi _{n}\right\rangle \right| ^{2}\cos \left( \omega
_{0n}t\right) =\left[ \hat{x}\left( 0\right) ,\hat{p}\left( t\right) \right]
, 
\]
that in the limit $t\rightarrow 0$ becomes an example of the well known
Thomas-Reiche-Kuhn sum rule. The rule is usually applied to atoms where a
sum over the three coordinates of the Z electrons is performed, so that it
reads 
\[
2m\sum_{n}\omega _{0n}\left| \left\langle \psi _{0}\left| \sum_{j=1}^{Z}%
\mathbf{r}_{j}\right| \psi _{n}\right\rangle \right|
^{2}=-i\sum_{k=1}^{3Z}\left[ \hat{x}_{k}\left( 0\right) ,\hat{p}_{k}\left(
0\right) \right] =3Z
\rlap{\protect\rule[1.1ex]{.325em}{.1ex}}h%
. 
\]

For a stationary process both the selfcorrelation and the commutator depend
only on the time difference $\left( t-t^{\prime }\right) .$ In this case the
latter eq.$\left( \ref{comm}\right) $ may be easily inverted via a time
integral. In fact we get 
\begin{eqnarray*}
&&\int_{-\infty }^{\infty }\sin \left[ \nu \left( t-t^{\prime }\right)
\right] \left[ x\left( t\right) ,x\left( t^{\prime }\right) \right] dt \\
&=&\int_{-\infty }^{\infty }\sin \left[ \nu \left( t-t^{\prime }\right)
\right] dt2i\int_{0}^{\infty }S_{x}\left( \omega \right) \sin \left[ \omega
\left( t^{\prime }-t\right) \right] d\omega \\
&=&-2i\int_{0}^{\infty }S_{x}\left( \omega \right) d\omega \int_{-\infty
}^{\infty }\sin \left[ \nu \left( t-t^{\prime }\right) \sin \left[ \omega
\left( t-t^{\prime }\right) \right] \right] dt=-i\pi S_{x}\left( \nu \right)
,
\end{eqnarray*}
where in the latter equality we take into account that $S_{x}\left( v\right)
=0$ for $v<0.$

All stationary properties of the SED oscillator studied in Section 2 may be
equally well obtained either from the spectrum eq.$\left( \ref{oscilspectrum}%
\right) ,$ from the selfcorrelation or from the commutator, the latter being 
\begin{eqnarray}
\left[ x\left( 0\right) ,x\left( t\right) \right] &=&2i 
\rlap{\protect\rule[1.1ex]{.325em}{.1ex}}h%
\int_{0}^{\infty }\frac{\tau \omega ^{3}\sin \left[ \omega t\right] d\omega 
}{\pi m\left[ \left( \omega _{0}^{2}-\omega ^{2}\right) ^{2}+\tau ^{2}\omega
^{6}\right] }  \nonumber \\
&=&i
\rlap{\protect\rule[1.1ex]{.325em}{.1ex}}h%
\int_{-\infty }^{\infty }\frac{\tau \omega ^{3}\exp \left[ i\omega t\right]
d\omega }{\pi m\left[ \left( \omega _{0}^{2}-\omega ^{2}\right) ^{2}+\tau
^{2}\omega ^{6}\right] }.  \label{com1}
\end{eqnarray}
This equality shows the advantage of the commutator with respect to the
selfcorrelation in the stochastic process associated to the SED oscillator.
In fact the latter integral may be performed analytically via the method of
residues, whilst getting the selfcorrelation requires approximations. 
\textit{Thus I propose that the reason for the use of commutators in QM is
the fact that the basic stochastic processes involved have spectra that are
odd with respect to the change }$\omega \rightarrow -\omega .$

Performing the integral eq.$\left( \ref{com1}\right) $ is now
straightforward. For $t>0$ we shall take into account the three simple poles
in the upper half plane of the complex variable $\omega $, that is 
\[
\omega =\pm \omega _{0}+\frac{1}{2}i\tau \omega _{0}^{2}+O\left( \tau
^{2}\omega _{0}^{3}\right) ,\omega =\pm i\left( \frac{1}{\tau }+\tau \omega
_{0}^{2}\right) +O\left( \tau ^{2}\omega _{0}^{3}\right) . 
\]
For $t<0$ we shall use the poles in the lower half plane. The contribution
of the poles in the imaginary axis should be neglected because it contains
an exponential of the form $\exp \left( -\left| t\right| /\tau \right) $
that is zero except for extremely small values of time. (Actually the term
derives from the high-frequency part of the spectrum and should be cut-off,
as discussed in Section 1.4.) Thus the result may be written, to order $%
O\left( \tau \omega _{0}^{2}\right) ,$%
\begin{equation}
\left[ x\left( 0\right) ,x\left( t\right) \right] =\frac{i 
\rlap{\protect\rule[1.1ex]{.325em}{.1ex}}h%
}{m\omega _{0}}\left\{ \sin \left( \omega _{0}t\right) +\tau \omega _{0}%
\frac{t}{\left| t\right| }\cos \left( \omega _{0}t\right) \right\} \exp
\left( -\frac{1}{2}\tau \omega _{0}^{2}\left| t\right| \right) ,  \label{xx}
\end{equation}
Similarly, taking eqs. $\left( \ref{comm}\right) $ and $\left( \ref{canonmom}%
\right) $ into account we may obtain the commutator of the canonical
momentum, that is 
\begin{equation}
\left[ p\left( 0\right) ,p\left( t\right) \right] =i 
\rlap{\protect\rule[1.1ex]{.325em}{.1ex}}h%
m\omega _{0}\sin \left( \omega _{0}t\right) \exp \left( -\frac{1}{2}\tau
\omega _{0}^{2}\left| t\right| \right) .  \label{pp}
\end{equation}
In the limit $\tau \rightarrow 0$ the commutator eq.$\left( \ref{xx}\right) $
agrees with the one derived from elementary quantum mechanics for the
corresponding (time dependent) operators in the Heisenberg picture.

The commutator of the particle coordinate in the SED oscillator may be
derived from the commutator of the electric field of the zeropoint radiation
taking eqs.$\left( \ref{Espectrum}\right) $ and $\left( \ref{comm}\right) $
into account. Actually the relation between the commutator of the vacuum
field and the commutator of a charged particle immersed in the field may be
obtained in the context of quantum mechanics without passing through the
spectrum. Indeed this was made long ago by Schiller for quantum commutators%
\cite{dice},\cite{Milonni}.

Our definition of commutator may be generalized to two different stationary
stochastic processes as follows:

\begin{definition}
Given two stationary stochastic process, $x(t)$\ and $y(t),$\ I define the
(stochastic) commutator of these processes, $\left[ x\left( t\right)
,y\left( t^{\prime }\right) \right] ,$\ as $2i$\ times the Hilbert transform
of the crosscorrelation, $\left\langle x\left( t\right) y\left( t^{\prime
}\right) \right\rangle .$
\end{definition}

The Hilbert transform, $g(u),$ of a function $f(t),t\in \left( -\infty
,\infty \right) $ is defined by 
\[
g(u)=\frac{1}{\pi }P\int_{-\infty }^{\infty }f(t)\frac{1}{u-t}dt,f\left(
t\right) =\frac{1}{\pi }P\int_{-\infty }^{\infty }g\left( u\right) \frac{1}{%
u-t}du, 
\]
where $P$\ means principal part and the second equality corresponds to the
inverse transform. However the inverse transform does not always recovers
the original. For instance the Hilbert transform of a constant is zero and
the inverse of zero is also zero. The relevant property for us is that the
Hilbert transforms changes sin$\left( \omega t\right) $ into cos$\left(
\omega u\right) $ and cos$\left( \omega t\right) $ into -sin$\left( \omega
u\right) ,$ provided that $\omega \neq 0$.

After that it is possible to define the derivative of a commutator with
respect to time, that is 
\begin{eqnarray*}
\frac{d}{dt}\left[ x\left( t\right) ,x\left( t^{\prime }\right) \right]
&=&\lim_{t^{\prime \prime }\rightarrow t^{\prime }}\frac{\left[ x\left(
t\right) ,x\left( t^{\prime \prime }\right) \right] -\left[ x\left( t\right)
,x\left( t^{\prime }\right) \right] }{t^{\prime \prime }-t^{\prime }} \\
&=&\lim_{t^{\prime \prime }\rightarrow t^{\prime }}\frac{\left[ x\left(
t\right) ,x\left( t^{\prime \prime }\right) -x\left( t^{\prime }\right)
\right] }{t^{\prime \prime }-t^{\prime }}=\left[ x\left( t\right) ,\frac{%
dx\left( t^{\prime }\right) }{dt^{\prime }}\right] ,
\end{eqnarray*}
where the linearity of the commutator has been used. Hence taking eq.$\left( 
\ref{xx}\right) $ into account we get to zeroth order in $\tau $%
\[
\left[ x\left( 0\right) ,p\left( t\right) \right] =i 
\rlap{\protect\rule[1.1ex]{.325em}{.1ex}}h%
\cos \left( \omega _{0}t\right) \Rightarrow \left[ x\left( 0\right) ,p\left(
0\right) \right] =i
\rlap{\protect\rule[1.1ex]{.325em}{.1ex}}h%
, 
\]
the latter being the fundamental commutation rule of quantum mechanics.

\textit{The stochastic commutator provides a hint for a realistic
interpretation of the quantum commutation rules as a disguised form of
stablishing the properties of some peculiar stochastic processes. The main
peculiarity is the fact that the spectra of the processes are usually odd
with respect to a change }$\omega \rightarrow -\omega $ \textit{of the
frequency}.

\section{Coupled oscillators in SED}

The generalization of the harmonic oscillator in SED to many dimensions is
straightforward using the appropriate extension of eq.$\left( \ref{ode}%
\right) .$ In the following I will study two simple examples of coupled
oscillators. Firstly a system of two one-dimensional oscillators at a long
distance as an example of van der Waals force. The system is interesting
because it shows that a phenomenon similar to quantum entanglement appears
also in SED. The second example is an array of coupled three-dimensional
oscillators at a finite temperature, that reproduces Debye theory of the
specific heat of solids.

\subsection{A model for quantum entanglement}

Entanglement is a quantum property of systems with several degrees of
freedom, which appears when the total state vector cannot be written as a
product of vectors associated to one degree of freedom each. In formal terms
a typical entangled state fulfils 
\begin{equation}
\mid \psi \left( 1,2\right) \rangle =\sum_{m,n}c_{mn}\mid \psi _{m}\left(
1\right) \rangle \mid \psi _{n}\left( 2\right) \rangle ,  \label{entangled}
\end{equation}
where $1$ and $2$ correspond to two different degrees of freedom, usually
belonging to different subsystems. The essential condition is that the state
eq.$\left( \ref{entangled}\right) $ cannot be written as a product, that is
the sum cannot be reduced to just one term via a change of basis in the
Hilbert space. Entanglement appears as a specifically quantum form of
correlation, which is claimed to be dramatically different from the
correlations of classical physics. The latter may be usually written in the
form 
\begin{equation}
\rho \left( 1,2\right) =\sum_{m,n}w_{mn}\rho _{m}\left( 1\right) \rho
_{n}\left( 2\right) ,  \label{ro12}
\end{equation}
where the quantities $\rho \gtrsim 0$ are probability densities and the
coefficients play the role of weights fulfilling $w_{mn}\gtrsim 0$, in sharp
contrast with eq.$\left( \ref{entangled}\right) $ where $\mid \psi \rangle $
are vectors in a Hilbert space and $c_{mn}$ are complex numbers.

In the last decades entanglement has been the subject of intense study, and
a resource for many applications, specially in the domain of quantum
information. In this case the relevant entanglement usually involves spin or
polarization. Entanglement is quite common in nonrelativistic quantum
mechanics of many-particle systems, e.g. for electrons in atoms or
molecules. However it is most relevant when the state-vectors $\mid \psi
_{m}\left( 1\right) \rangle $ and $\mid \psi _{n}\left( 2\right) \rangle $
of eq.$\left( \ref{entangled}\right) $ belong to different systems placed
far from each other. A study of entanglement and its relation with ``local
realism'' will be made in Chapter 4 and examples of photon entanglement will
be provided in Chapter 6. Here I will illustrate, with a simple example,
that entanglement might be understood as a correlation induced by quantum
vacuum fluctuations acting in two different places.

\subsection{London-van der Waals forces}

I shall study the London theory of the van der Waals forces in a simple
model of two one-dimensional oscillating electric dipoles. Each dipole
consists of a particle at rest and another particle (which we will name
electron) with mass $m$ and charge $e$. In the model it is assumed that
every electron moves in a harmonic oscillator potential and there is an
additional interaction between the electrons. Thus the Hamiltonian is 
\begin{equation}
H=\frac{p_{1}^{2}}{2m}+\frac{1}{2}m\omega _{0}^{2}x_{1}^{2}+\frac{p_{2}^{2}}{%
2m}+\frac{1}{2}m\omega _{0}^{2}x_{2}^{2}-Kx_{1}x_{2},  \label{dip}
\end{equation}
where $x_{1}(x_{2})$ is the position of the electron of the first (second)
dipole with respect to the equilibrium position. The positive parameter $%
K<m\omega _{0}^{2}$ depends on the distance bewteen the dipoles, but the
dependence is irrelevant for our purposes. (For a more complete study of
this problem within SED see Refs.\cite{dice}, \cite{dice2}). We shall work
both the QM and the SED calculations.

\subsection{Quantum theory of the model}

An exact quantum calculation is not difficult. We take $x_{j},p_{j}$ and $H$
as operators in the Hilbert space of the full system, fulfilling the
standard commutation relations 
\begin{equation}
\left[ \hat{x}_{j},\hat{x}_{l}\right] =\left[ \hat{p}_{j},\hat{p}_{l}\right]
=0,\left[ \hat{x}_{j},\hat{p}_{l}\right] =i
\rlap{\protect\rule[1.1ex]{.325em}{.1ex}}h%
\delta _{jl}.  \label{7.10}
\end{equation}
Now we introduce the new operators 
\begin{eqnarray}
\hat{x}_{+}\left( t\right) &=&\frac{1}{\sqrt{2}}\left[ \hat{x}_{1}\left(
t\right) +\hat{x}_{2}\left( t\right) \right] ,\hat{x}_{-}\left( t\right) =%
\frac{1}{\sqrt{2}}\left[ \hat{x}_{1}\left( t\right) -\hat{x}_{2}\left(
t\right) \right] ,  \nonumber \\
p_{+}\left( t\right) &=&\frac{1}{\sqrt{2}}\left[ \hat{p}_{1}\left( t\right) +%
\hat{p}_{2}\left( t\right) \right] ,\hat{p}_{-}\left( t\right) =\frac{1}{%
\sqrt{2}}\left[ \hat{p}_{1}\left( t\right) -\hat{p}_{2}\left( t\right)
\right] .  \label{7.11}
\end{eqnarray}
It is easy to derive the commutation relations of the new operators, that
are similar to eqs.$\left( \ref{7.10}\right) $ with the subindices $+,-$
susbstituted for $1,2$. The Hamiltonian eq.$\left( \ref{dip}\right) $ in
terms of the new operatos is 
\[
\hat{H}=\frac{\hat{p}_{+}^{2}}{2m}+\frac{1}{2}\left( m\omega
_{0}^{2}+K\right) \hat{x}_{+}^{2}+\frac{\hat{p}_{-}^{2}}{2m}+\frac{1}{2}%
\left( m\omega _{0}^{2}-K\right) \hat{x}_{-}^{2}. 
\]
This is equivalent to two uncoupled harmonic oscillators with the same mass, 
$m$, and frequencies 
\[
\omega _{+}=\sqrt{\omega _{0}^{2}+K/m},\omega _{-}=\sqrt{\omega _{0}^{2}-K/m}%
,\left( K<m\omega _{0}^{2}\right) 
\]
respectively$.$ Thus the wavefunction of the two-electron system is 
\begin{eqnarray}
\psi &=&\psi \left( x_{+}\right) \psi \left( x_{-}\right) =\sqrt{\frac{m}{%
\pi 
\rlap{\protect\rule[1.1ex]{.325em}{.1ex}}h%
\sqrt{\omega _{+}\omega _{-}}}}\exp \left[ -\frac{m}{2 
\rlap{\protect\rule[1.1ex]{.325em}{.1ex}}h%
}\left( \omega _{+}x_{+}^{2}+\omega _{-}x_{-}^{2}\right) \right]
\label{7.12} \\
&=&\sqrt{\frac{m}{\pi 
\rlap{\protect\rule[1.1ex]{.325em}{.1ex}}h%
\sqrt{\omega _{+}\omega _{-}}}}\exp \left\{ -\frac{m}{4 
\rlap{\protect\rule[1.1ex]{.325em}{.1ex}}h%
}\left[ (\omega _{+}+\omega _{-})\left( x_{1}^{2}+x_{2}^{2}\right) +2(\omega
_{+}-\omega _{-})\left( x_{1}x_{2}\right) \right] \right\} ,  \nonumber
\end{eqnarray}
and the interaction energy of the system is 
\[
\Delta E=\frac{
\rlap{\protect\rule[1.1ex]{.325em}{.1ex}}h%
}{2}\left( \sqrt{\omega _{0}^{2}-K/m}+\sqrt{\omega _{0}^{2}-K/m}-2\omega
_{0}\right) =-\frac{
\rlap{\protect\rule[1.1ex]{.325em}{.1ex}}h%
K^{2}}{4m^{2}\omega _{0}^{3}}+O\left( K^{4}\right) , 
\]
that to lowest nontrivial order in the coupling constant $K$ gives 
\begin{equation}
\psi =\sqrt{\frac{m\omega _{0}}{\pi 
\rlap{\protect\rule[1.1ex]{.325em}{.1ex}}h%
}}\left( 1+\frac{2Kx_{1}x_{2}}{m\omega _{0}}\right) \exp \left[ -\frac{%
m\omega _{0}}{2
\rlap{\protect\rule[1.1ex]{.325em}{.1ex}}h%
}\left( x_{1}^{2}+x_{2}^{2}\right) \right] ,  \label{7.12a}
\end{equation}
which may be written in terms of the wavefunctions of the ground state, $%
\psi _{0}\left( x\right) ,$ and the first excited state, $\psi _{1}\left(
x\right) ,$ of the simple oscillator as follows 
\[
\psi =\psi _{0}\left( x_{1}\right) \psi _{0}\left( x_{2}\right) +\frac{K}{%
m\omega _{0}^{2}}\psi _{1}\left( x_{1}\right) \psi _{1}\left( x_{2}\right) . 
\]
(The function is not normalized because the normalization was lost when we
truncated at first order the expansion in powers of $K$). In quantum
language this wavefunction $\psi $ may be interpreted saying that the
two-system state is a sum of two amplitudes, one of them corresponds to both
oscillators being in the ground state and the other one to both being in the
first excited state. It is true that eq.$\left( \ref{7.12}\right) $ is not
an irreducible sum of products like eq.$\left( \ref{entangled}\right) .$
However, in cannot be factorized in terms of wavefunctions of individual
electrons and therefore it is not a classical correlation that might be
represented as eq.$\left( \ref{ro12}\right) $. Therefore it may by
considered an entangled state involving two distant systems.

Although quantum mechanics usually does not offer intuitive pictures of the
phenomena, in this case it is difficult to refrain from interpreting the
entanglement in this example as a correlation of the (random) motions of the
electrons. Indeed the modulus squared of the wavefunction eq.$\left( \ref
{7.12}\right) $ gives the probability density for the positions of the
electrons, which is larger when the electrons are far from each other so
that their mutual repulsion energy is smaller. Then the correlation
(entanglement) lowers the energy giving rise to an attractive force between
the oscillators. Of course this explanation departs from the Copenhagen
interpretation (see Chapter 3), that should not speak about the probability
that \emph{one electron is} in the region $x_{1}>0$ and the \emph{other one} 
\emph{is} in the region $x_{2}>0$. Instead it compels us to say something
like ``if we perform a measurement of the simultaneous positions of the
electrons \emph{the probability that we get} one of them in the region $%
x_{1}>0$ and the other one is in the region $x_{2}>0$ is given by the
modulus squared of eq.$\left( \ref{7.12}\right) "$. (Simultaneous
measurements are possible because the observables commute.) In any case the
origin of the correlation is not clear in quantum mechanics.

\subsection{The model in stochastic electrodynamics}

In sharp contrast with QM the interpretation offered by SED is transparent:
the random motion of the electrons is induced by the ZPF, and the
correlation is produced by the interaction. The SED calculation is as
follows. The differential equations of motion may be obtained from eq.$%
\left( \ref{dip}\right) $. I shall write them including the forces due to
the random ZPF and the radiation reaction, see eq.$\left( \ref{ode}\right) ,$
that is 
\begin{eqnarray}
m\stackrel{..}{x_{1}} &=&-m\omega _{0}^{2}x_{1}-Kx_{2}+\frac{2e^{2}}{3c^{3}}%
\stackrel{...}{x_{1}}+eE_{1}\left( t\right) ,  \nonumber \\
m\stackrel{..}{x}_{2} &=&-m\omega _{0}^{2}x_{2}-Kx_{1}+\frac{2e^{2}}{3c^{3}}%
\stackrel{...}{x_{2}}+eE_{2}\left( t\right) .  \label{ode2}
\end{eqnarray}
The approximation of neglecting the $x$ dependence of the field, $E(\mathbf{%
x,}t)$, is not good if the dipoles are at a long distance (on the other hand
the Hamiltonian eq.$\left( \ref{dip}\right) $ is not valid for short
distances). However we may neglect the $x$ dependence within each dipole,
that is we will approximate $E\left( \mathbf{x}_{1,}t\right) \simeq $ $%
E\left( \mathbf{a},t\right) ,E\left( \mathbf{x}_{2},t\right) \simeq $ $%
E\left( \mathbf{b,}t\right) ,$ where $\mathbf{a}$ and $\mathbf{b}$ are the
positions of the first and second dipole, respectively. Also we will
simplify the notation writing $E_{1}\left( t\right) $ for $E\left( \mathbf{a,%
}t\right) $ and $E_{2}\left( t\right) $ for $E\left( \mathbf{a,}t\right) .$
Furthermore, as we assume that the distance between dipoles is large, we
shall take the stochastic processes $E_{1}\left( t\right) $ and $E_{2}\left(
t\right) $ as uncorrelated.

The coupled eqs.$\left( \ref{ode2}\right) $ may be decoupled via writing new
equations which are the sum and the difference of the former, and
introducing the new position variables 
\begin{equation}
x_{+}\left( t\right) =\frac{1}{\sqrt{2}}\left[ x_{1}\left( t\right)
+x_{2}\left( t\right) \right] ,x_{-}\left( t\right) =\frac{1}{\sqrt{2}}%
\left[ x_{1}\left( t\right) -x_{2}\left( t\right) \right] ,  \label{ode5}
\end{equation}
and similarly definitions for $E_{+}\left( t\right) $ and $E_{-}\left(
t\right) .$ We get 
\begin{eqnarray}
m\stackrel{..}{x_{+}} &=&-(m\omega _{0}^{2}-K)x_{+}+\frac{2e^{2}}{3c^{3}}%
\stackrel{...}{x_{+}}+eE_{+}\left( t\right) ,  \nonumber \\
m\stackrel{..}{x}_{-} &=&-(m\omega _{0}^{2}+K)x_{-}+\frac{2e^{2}}{3c^{3}}%
\stackrel{...}{x_{-}}+eE_{-}\left( t\right) ,  \label{ode3}
\end{eqnarray}
where the stochastic processes $E_{+}\left( t\right) $ and $E_{-}\left(
t\right) $ are statistically independent as a consequence of $E_{1}\left(
t\right) $ and $E_{2}\left( t\right) $ being uncorrelated. With the method
used to solve eqs.$\left( \ref{2.3}\right) $ and $\left( \ref{2.4}\right) $
we get 
\begin{equation}
\left\langle x_{\pm }^{2}\right\rangle =\frac{
\rlap{\protect\rule[1.1ex]{.325em}{.1ex}}h%
}{2m\sqrt{\omega _{0}^{2}\mp K/m}},\left\langle v_{\pm }^{2}\right\rangle =%
\frac{
\rlap{\protect\rule[1.1ex]{.325em}{.1ex}}h%
\sqrt{\omega _{0}^{2}\mp K/m}}{2m}.  \label{xv}
\end{equation}
The Hamiltonian eq.$\left( \ref{dip}\right) $ may be written in terms of $%
x_{+}\left( t\right) $, $x_{-}\left( t\right) $ leading to 
\[
H=\frac{p_{+}^{2}}{2m}+\frac{1}{2}m\omega _{0}^{2}x_{+}^{2}+\frac{p_{-}^{2}}{%
2m}+\frac{1}{2}m\omega _{0}^{2}x_{-}^{2}-\frac{1}{2}K\left(
x_{+}^{2}-x_{-}^{2}\right) . 
\]
Hence, defining $p_{\pm }=mv_{\pm },$ it is easy to get the total energy, $%
\left\langle H\right\rangle ,$ taking eqs.$\left( \ref{xv}\right) $ into
account. The result is in agreement with the quantum eq.$\left( \ref{7.13}%
\right) .$ The joint probability distribution of positions is Gaussian and
factorizes because eqs.$\left( \ref{ode3}\right) $ are decoupled. That is 
\[
\rho \left( x_{+},x_{-}\right) dx_{+}dx_{-}=\rho _{+}\left( x_{+}\right)
\rho _{-}\left( x_{-}\right) dx_{+}dx_{-}. 
\]
The densities $\rho _{\pm }$ should be normalized whence we get 
\[
\rho _{\pm }\left( x\right) =\sqrt{\frac{2m}{\pi 
\rlap{\protect\rule[1.1ex]{.325em}{.1ex}}h%
}}\left( \omega _{0}^{2}\mp K/m\right) ^{-1/4}\exp \left[ -\frac{m}{2 
\rlap{\protect\rule[1.1ex]{.325em}{.1ex}}h%
}\sqrt{\omega _{0}^{2}\mp K/m}x_{\pm }^{2}\right] . 
\]
Hence it is easy to get the joint probability in terms of the variables $%
x_{1}$ and $x_{2}$ taking eqs.$\left( \ref{ode5}\right) $ into account. The
result is in agreement with the quantum prediction, eq.$\left( \ref{7.12}%
\right) .$

In the equation of motion$\left( \ref{ode2}\right) $ I have assumed that the
ZPF components, $E_{1}\left( t\right) $ and $E_{2}\left( t\right) $, acting
upon the two particles are uncorrelated. This is a good approximation if the
particles are at a distance which is large in comparison with wave lenght, $%
\lambda \simeq c/\omega _{0}$, corresponding to the typical frequencies
involved. However if the distance is of that order or smaller, the ZPF
components will be correlated, which would cause a much stronger correlation
between the particle$\acute{}$s motions. We might speculate that
correlations induced by the ZPF are related to quantum statistics, that is
behaviour of particles as either bosons or fermions. But this possibility
will not be further discussed in this book.

\textit{The conclusion of our study of the two coupled oscillators in SED is
the suggestion that quantum entanglement is a correlation between the
quantum fluctuations of different systems, mediated by the vacuum fields,
these fields not being apparent in the quantum formalism.}

\subsection{Specific heats of solids}

An application of SED at a finite temperature is the calculation of the
specific heat of solids, which we summarize in the following\cite{Blanco2}.
We shall consider a solid as a set of positive ions immersed in an electron
gas. As is well known the electrons contribute but slightly to the specific
heat at not too high temperatures. In SED we shall study the motion of the
ions under the action of three forces. The first one derives from the
interaction with the neighbour ions and the electron gas, that may be
modelled by an oscillator potential which increases when the distance
between neighbour ions departs from the equilibrium configuration. The
second is the random background radiation with Planck spectrum (including
the ZPF) and the third one is the radiation reaction. This gives rise to a
discrete set of coupled third order differential equations that may be
decoupled by the introduction of normal mode coordinates. After that, every
equation is similar to eq.$\left( \ref{ode}\right) $ and may be solved in
anologous form. The net result is that the mean (potential plus kinetic)
energy in equilibrium becomes 
\begin{equation}
E(\omega )=\frac{1}{2}
\rlap{\protect\rule[1.1ex]{.325em}{.1ex}}h%
\omega \coth \left( \frac{
\rlap{\protect\rule[1.1ex]{.325em}{.1ex}}h%
\omega }{2kT}\right) ,  \label{solids}
\end{equation}
where $\omega $ is the frequency of the mode. With an appropriate
distribution, $\rho \left( \omega \right) ,$ of modes this leads to the
quantum result derived by Debye\cite{Debye12}, the specific heat being the
derivative of the total energy with respect to the temperature.

There are other interesting results of SED at a finite temperature, in
particular about magnetic properties. They may be seen in the books of de la
Pe\~{n}a et al.\cite{dice}, \cite{dice2} and references therein.

\textit{The SED calculation of the specific heat of solids provides another
argument for the continuity (as opposed to discreteness) of the energies of
quantum oscillators. If it is hard to accept that electromagnetic radiation
consists of particles (photons) in a realistic interpretation of quantum
physics, it is still harder to assume that quantized oscillations of the
ions in a solid ( phonons) are particles. It is more plausible to assume
that the energies of the normal modes of the set of ions have a continuous,
although random, distribution of energies such that the average for a mode
is given by eq.}$\left( \ref{solids}\right) .$\textit{\ Also it is plausible
that the mean energy of a vibration mode of every ion is the same as the
mean energy of the radiation mode having the same frequency, that is the
result here obtained.}

\section{The particle in a homogeneous magnetic field}

Another linear problem that has been extensively studied within SED is the
motion of a charged particle in a homogeneous magnetic field\cite{dice}. The
most relevant result is the prediction of diamagnetic properties of a free
charge (without magnetic moment), which departs from classical physics and
agrees with QM. Here I shall revisit the SED calculation of the free charged
particle in a homogeneous field of magnitude $B$.

\subsection{Classical theory}

The classical motion may be got from Newton\'{}s law\textbf{\ }with the
Lorentz force, that is 
\begin{equation}
m\stackrel{..}{\mathbf{r}}=\mathbf{-}\left( e/c\right) \stackrel{\cdot }{%
\mathbf{r}}\times \mathbf{B}.  \label{Lorentzforce}
\end{equation}
\textbf{\ }If we choose the Z axis in the direction of the $\mathbf{B}$ the
motion in that direction is uniform and in the perpendicular plane it is
given by 
\begin{eqnarray}
x &=&R\cos \left[ 2\omega _{0}\left( t-t_{0}\right) \right] +x_{0}, 
\nonumber \\
y &=&R\sin \left[ 2\omega _{0}\left( t-t_{0}\right) \right]
+y_{0},\smallskip \omega _{0}\equiv \frac{eB}{2mc},  \label{classmag}
\end{eqnarray}
with four integration constants, namely $\{R,t_{0},x_{0},x_{0}\}.$ The
motion is circular with radius $R$ and constant (Larmor) angular frequency $%
\omega _{0}.$ The total energy $E$ may be identified with the Hamiltonian,
that is 
\begin{equation}
H=\frac{p_{x}^{2}}{2m}+\frac{p_{y}^{2}}{2m}-\omega _{0}\left(
xp_{y}-yp_{x}\right) +\frac{1}{2}m\omega _{0}^{2}\left( x^{2}+y^{2}\right) .
\label{Hmag}
\end{equation}
Taking Hamilton equations into account we get 
\begin{equation}
E=\frac{1}{2}m\left( \stackrel{\cdot }{x}^{2}+\stackrel{\cdot }{y}%
^{2}\right) =2mR^{2}\omega _{0}^{2}.  \label{magenergy}
\end{equation}

Actually in a classical electrodynamical calculation we should include the
radiation reaction (similar to the second term of the right side in the
oscillator eq.$\left( \ref{ode}\right) ).$ This term would give rise to a
loss of energy by radiation whence the system will eventually arrive at the
state of minimal energy, that is zero (when $R=0$). This shows that no
diamagnetic effects can be expected to occur in classical physics.

\subsection{Quantum theory}

The QM treatment starts from a quantum Hamiltonian operator which may be got
from eq.$\left( \ref{Hmag}\right) $ by promoting the classical coordinates
and momenta to operators in a Hilbert space (for a detailed study see\cite
{Dicke}). The Z component of the angular momentum operator and the
Hamiltonian commute and we may search for simultaneous eigenvectors having
eigenvalues 
\begin{eqnarray}
L_{z} &\equiv &xp_{y}-yp_{x}\rightarrow m_{l}
\rlap{\protect\rule[1.1ex]{.325em}{.1ex}}h%
,m_{l}=0,\pm 1,\pm 2,...,  \nonumber \\
H &\rightarrow &E_{r}=\left( 2r+1\right) 
\rlap{\protect\rule[1.1ex]{.325em}{.1ex}}h%
\left| \omega _{0}\right| ,r=0,1,2,...  \label{Qmag}
\end{eqnarray}
We see that the quantum ground state, given by $r=0$ and $E_{r}= 
\rlap{\protect\rule[1.1ex]{.325em}{.1ex}}h%
\left| \omega _{0}\right| $, has an infinite degeneracy because this energy
is shared by states with all possible values of $m_{l}.$ For this reason it
is common to add to the Hamiltonian a two-dimensional oscillator potential
with characteristic frequency $\omega _{1}(>0.)$ Then the energy eigenvalues
have an additional term $\left( 2n+1\right) 
\rlap{\protect\rule[1.1ex]{.325em}{.1ex}}h%
\omega _{1}$ with $n=2r-m_{l}\geq 0,$which breaks the degeneracy, the ground
state now corresponding to $r=n=m_{l}=0.$ From eq.$\left( \ref{Qmag}\right) $
we may get the most relevant parameter, that is the magnetic moment. In the
ground state it is 
\begin{equation}
\mathbf{M}=-\nabla _{\mathbf{B}}E=-
\rlap{\protect\rule[1.1ex]{.325em}{.1ex}}h%
\nabla _{\mathbf{B}}\left| \omega _{0}\right| =-M_{B}\frac{\mathbf{B}}{B}%
,M_{B}=\frac{
\rlap{\protect\rule[1.1ex]{.325em}{.1ex}}h%
\left| e\right| }{2mc}  \label{magmom}
\end{equation}
where $M_{B}$ is the Bohr magneton. This (or the ground state energy, second
eq.$\left( \ref{Qmag}\right) )$ is the result that we may expect to
reproduce in SED.

\subsection{SED treatment}

In SED we should add the action of the ZPF (plus the radiation reaction) to
the force derived from the homogeneous magnetic field, see eq.$\left( \ref
{Lorentzforce}\right) $. We will study only the motion in the $XY$ plane. If 
$u\equiv $ $\stackrel{\cdot }{x}$ and $v$ $\equiv $ $\stackrel{\cdot }{y}$
are the components of the velocity vector the equations of motion are 
\begin{equation}
m\stackrel{\cdot }{u}=\frac{e}{c}vB+m\tau \ddot{u}+eE_{u},\text{ }m\stackrel{%
\cdot }{v}=-\frac{e}{c}uB+m\tau \stackrel{\cdot \cdot }{v}+eE_{v},
\label{magn}
\end{equation}
where the first term is the component of the Lorentz force, the second is
the radiation reaction and the third one the action of the ZPF (in the long
wavelength approximation, see eq.$\left( \ref{ode}\right) ).$ The components
of the electric ZPF, $E_{u}\left( t\right) $ and $E_{v}\left( t\right) ,$
are assumed statistically independent stochastic processes.

The small value of $\tau <<1/\omega _{0}$ allows an approximation similar to
the one made in the free particle case, Section 3.4. We may substitute $%
evB/(cm\tau )$ for $\ddot{u}$ and similar for $\stackrel{\cdot \cdot }{v},$
thus obtaining two first order equations from eqs.$\left( \ref{magn}\right)
. $ Then the solution is straightforward and we get, with steps similar to
those involved in the solution of eq.$\left( \ref{ode}\right) $ lead to 
\begin{equation}
\left\langle u^{2}\right\rangle =\int_{0}^{\infty }\frac{%
\rlap{\protect\rule[1.1ex]{.325em}{.1ex}}h%
\tau \omega ^{3}\left( 4\omega _{0}^{2}+\omega ^{2}\right) }{\pi m\left[
\left( 4\omega _{0}^{2}-\omega ^{2}\right) ^{2}+4\tau ^{2}\omega ^{6}\right] 
}d\omega \simeq \frac{
\rlap{\protect\rule[1.1ex]{.325em}{.1ex}}h%
\left| \omega _{0}\right| }{m},  \label{mag2}
\end{equation}
and the same result for $\left\langle v^{2}\right\rangle .$ Actually the
integral in eq.$\left( \ref{mag2}\right) $ is ultraviolet divergent so that
a a high frequency cutoff, $\omega _{c}$, should be included. It may be seen
that for small $\tau $, i. e. $\tau \omega _{0}<<1,$ the main contributions
to the integral eq.$\left( \ref{mag2}\right) $ come either from frequencies $%
\omega $ close to $2\omega _{0}$ or for high frequencies $\omega >>2\omega
_{0}$ (a similar case happens in the oscillator, see eqs.$\left( \ref{2.3}%
\right) $ and $\left( \ref{2.4}\right) $). The former contribution, given by
eq.$\left( \ref{mag2}\right) ,$ is independent of both the cut-off frequency
and the precise value of $\tau .$ The latter, high frequencies, contribution
may be obtained neglecting $\omega _{0}$ in comparison with $\omega ,$ and
putting $4\omega _{0}$ as lower limit of the integral in order to exclude
the frequency region around $2\omega _{0}$ calculated in eq.$\left( \ref
{mag2}\right) .$ Thus we get 
\begin{eqnarray*}
\left\langle u^{2}\right\rangle _{hf} &\simeq &\int_{4\omega _{0}}^{\omega
_{c}}\frac{
\rlap{\protect\rule[1.1ex]{.325em}{.1ex}}h%
\tau \omega ^{5}d\omega }{\pi m\left( \omega ^{4}+4\tau ^{2}\omega
^{6}\right) } \\
&=&\frac{
\rlap{\protect\rule[1.1ex]{.325em}{.1ex}}h%
}{8\pi m\tau }\log \left( 1+4\tau ^{2}\omega _{c}^{2}\right) \simeq \frac{%
\rlap{\protect\rule[1.1ex]{.325em}{.1ex}}h%
\tau }{2\pi m}\omega _{c}^{2},
\end{eqnarray*}
where we have assumed $2\tau \omega _{c}<<1$. A comparison with eq.$\left( 
\ref{4.10a}\right) $ shows that this contribution is the same for a free
particle. Indeed it is independent of the magnetic field (which does appear
in eq.$\left( \ref{mag2}\right) .$

The mean energy in SED is obtained inserting eq.$\left( \ref{mag2}\right) $
in the expression of the energy (see eq.$\left( \ref{magenergy}\right) )$
giving 
\begin{equation}
\left\langle E\right\rangle =\frac{1}{2}m\left\langle
u^{2}+v^{2}\right\rangle =
\rlap{\protect\rule[1.1ex]{.325em}{.1ex}}h%
\left| \omega _{0}\right| ,  \label{7.0}
\end{equation}
in agreement with the quantum result. Hence there is also agreement for the
magnetic moment, eq.$\left( \ref{magmom}\right) $\textrm{.}

Another interesting result from SED is the mean value of the angular
momentum that is 
\begin{equation}
\left\langle L_{z}\right\rangle =\left\langle xp_{y}-yp_{x}\right\rangle
=m\left\langle xv-yu\right\rangle =-\frac{e}{\left| e\right| } 
\rlap{\protect\rule[1.1ex]{.325em}{.1ex}}h%
,  \label{7.1a}
\end{equation}
independently of the magnitude of the magnetic field and the mass of the
particle. I omit the proof that is straightforward. Thus the angular
momentum is parallel to the magnetic field if the charge is negative and
antiparallel if it is positive. We saw that the magnetic moment is always
antiparallel to the magnetic field. The results eqs.$\left( \ref{7.0}\right) 
$ and $\left( \ref{7.1a}\right) $ correspond to the limit $\tau \rightarrow
0.$ In both cases there are corrections for finite $\tau $ which would
requiere a relativistic treatment. If an appropriate cutoff is introduced,
say $\omega _{c}=mc^{2}/
\rlap{\protect\rule[1.1ex]{.325em}{.1ex}}h%
,$ the high frequencies contribution is small.

There is however a disagreement between QM and SED for the angular momentum
in the stationary state. SED predicts a finite value given by eq.$\left( \ref
{7.1a}\right) $ but in QM there are many possible angular momenta in the
ground state as shown in eq.$\left( \ref{Qmag}\right) .$ On the other hand
if we include an additional oscillator potential with characteristic
frequency $\omega _{1}$, then the quantum prediction for the ground state
angular momentum is zero that also disagrees with the SED result. Thus in QM
there are two features whose realistic interpretation is difficult. Firstly
eq.$\left( \ref{magmom}\right) $ that strongly suggests that the angular
momentum in the ground state is the same of the SED prediction, eq.$\left( 
\ref{7.1a}\right) $ rather than the degeneracy eq.$\left( \ref{Qmag}\right)
. $ Secondly that an additional oscillator potential no matter how small
breaks the degeneracy, but leading to zero angular momentum, rather than the
most intuitive value eq.$\left( \ref{7.1a}\right) .$ These facts show that a
realistic interpretation of the angular momentum in quantum mechanics is
difficult. A possible solution is proposed in Section 6.4.

\textit{In summary the SED treatment of the particle in a homogeneous
magnetic field reproduces the most relevant results of QM and provides a
realistic interpretation for the QM prediction of a diamagnetic behaviour of
the charged particle in the presence of an homogeneous magnetic field.
However there is disagreement for the angular momentum.}

\section{SED application to nonlinear systems}

Several nonlinear systems have been studied in stochastic electrodynamics
that provide some results in semiquantitative agreement with quantum
mechanics, but badly fail in other cases. Actually SED reproduces quantum
results, and agrees with experiments, in a limited domain, namely for
systems of charged particles that may be treated linearly and within a
nonrelativistic approximation. In sharp contrast the treatment of nonlinear
systems gives results that frequently disagree with quantum predictions. The
explanation of this fact is that SED, as defined in the introduction
section, is an approximation to QED to lowest order in Planck constant $
\rlap{\protect\rule[1.1ex]{.325em}{.1ex}}h%
,$ but quantum mechanics gives predictions for nonlinear systems that
involve $
\rlap{\protect\rule[1.1ex]{.325em}{.1ex}}h%
$ to higher order. Therefore to be valid for all physical systems, SED
should be generalized, likely including all vacuum fields and taking into
account the back action of the particles on the fields. I propose that this
would lead to quantum theory, but the steps needed are not known. There is
here a paradox, namely we might foresee that the final theory should be
rather cumbersome due to the large number of fields involved and the
nonlinearity of equations, but it is the case that quantum theory has a
relatively simple formalism. This is the magic of quantum theory and one of
the reasons for the difficulty of getting a realistic interpretation of it.

In the following I comment on some calculations for nonliner systems . The
best method for the SED study in these cases is to get the evolution of the
classical mechanical `constants of the motion', one of them being the total
energy. These parameters are no longer constant due to the interaction with
the ZPF and the radiation reaction, but may be slowly varying. The method
was used in Section 3 for the oscillator and it will be illustrated in the
following for a nonlinear system, the rigid planar rotor. After that I will
comment on the hydrogen atom in SED and the problem of equilibrium between
radiation and matter.

\subsection{The planar rigid rotator}

The planar rigid rotor is the most simple nonlinear system studied in SED%
\cite{Boyerrotor}. A model of the rotor is a particle of mass $m$ and charge 
$e$ constrained to move in the $XY$ plane always at a distance $R$ of a
fixed point. Thus the problem has a single degree of freedom and the SED
equation of motion may be written in terms of the polar angle $\phi $ as
follows 
\begin{eqnarray*}
mR\stackrel{..}{\phi } &=&-m\tau R\stackrel{\cdot }{\phi }^{3}+m\tau 
\stackrel{...}{\phi }+eE, \\
E &=&-\cos \phi E_{x}\left( t\right) +\sin \phi E_{y}\left( t\right) .
\end{eqnarray*}
The former two terms of the right side give the tangential component of the
radiation reaction force, $m\tau \stackrel{...}{\mathbf{r}},$ and the
tangential component of the force due to the ZPF, respectively. In terms of
the angular velocity, $\omega =\stackrel{\cdot }{\phi },$ the equation
becomes 
\begin{mathletters}
\begin{equation}
\stackrel{\cdot }{\omega }=-\tau \omega ^{3}+\tau \stackrel{..}{\omega }+%
\frac{e}{mR}E.  \label{rotor}
\end{equation}
Eq.$\left( \ref{rotor}\right) $ may be solved perturbatively in two steps.
In the first step we solve the classical equation of motion $\stackrel{\cdot 
}{\omega }=0,$ which trivially gives $\omega =\omega _{0}=$constant. That
constant becomes slowly varying when we take into account the radiation
raction and the action of the ZPF. In order to get that variation eq.$\left( 
\ref{rotor}\right) $ may be solved substituting $\omega _{0}$ for $\omega $
in the perturbation, that is in all terms of the right side. The solution
with initial condition $\omega \left( 0\right) =\omega _{0}$ becomes 
\end{mathletters}
\[
\omega (t)=\omega _{0}+\int_{0}^{t}dt^{\prime }\left[ -\tau \omega _{0}^{3}+%
\frac{e}{mR}\left( -\cos \left( \omega _{0}t^{\prime }\right) E_{x}\left(
t\right) +\sin \left( \omega _{0}t^{\prime }\right) E_{y}\left( t^{\prime
}\right) \right) \right] 
\]
The former term within the integral sign represents drift and the latter
term diffusion. The diffusion constant may be calculated via the limit 
\begin{eqnarray}
D &=&\lim_{t\rightarrow \infty }\frac{1}{t}\left\langle \left[
\int_{0}^{t}dt^{\prime }\frac{e}{mR}\left( -\cos \left( \omega _{0}t^{\prime
}\right) E_{x}\left( t^{\prime }\right) +\sin \left( \omega _{0}t^{\prime
}\right) E_{y}\left( t^{\prime }\right) \right) \right] ^{2}\right\rangle 
\nonumber \\
&=&\left( \frac{e}{mR}\right) ^{2}\lim_{t\rightarrow \infty }\frac{1}{t}%
\int_{0}^{t}dt^{\prime }\int_{0}^{t}dt^{^{\prime \prime }}\cos \left[ \omega
_{0}\left( t^{\prime }-t^{\prime \prime }\right) \right] \left\langle
E_{x}\left( t^{\prime }\right) E_{x}\left( t^{\prime \prime }\right)
\right\rangle ,  \label{diffusion}
\end{eqnarray}
where I have taken into account that $\left\langle E_{x}\left( t^{\prime
}\right) E_{x}\left( t^{\prime \prime }\right) \right\rangle =\left\langle
E_{y}\left( t^{\prime }\right) E_{y}\left( t^{\prime \prime }\right)
\right\rangle $ and that $\left\langle E_{x}\left( t^{\prime }\right)
E_{y}\left( t^{\prime \prime }\right) \right\rangle =0.$ The field
correlation may be easily got from the spectrum, eq.$\left( \ref{Espectrum}%
\right) $ as follows 
\[
\left\langle E_{x}\left( t^{\prime }\right) E_{x}\left( t^{\prime \prime
}\right) \right\rangle =\frac{2
\rlap{\protect\rule[1.1ex]{.325em}{.1ex}}h%
}{3\pi c^{3}}\int_{0}^{\infty }u^{3}\cos \left[ u\left( t^{\prime
}-t^{\prime \prime }\right) \right] du. 
\]
When this is inserted in eq.$\left( \ref{diffusion}\right) $ the variables $%
t^{\prime }$ and $t^{\prime \prime }$ may be changed to $w\equiv \left(
t^{\prime }+t^{\prime \prime }\right) /2$ and $t^{\prime }-t^{\prime \prime
}\equiv s$ . With good approximation the integration may be performed from $%
0 $ to $t$ for the $w$ integral and for the whole real line for the variable 
$s $. Then the limit $t\rightarrow \infty $ in eq.$\left( \ref{diffusion}%
\right) $ is trivial and we get, taking the definition of $\tau ,$ eq.$%
\left( \ref{gamma}\right) ,$ into account, 
\begin{eqnarray*}
D &=&\frac{
\rlap{\protect\rule[1.1ex]{.325em}{.1ex}}h%
\tau }{\pi mR^{2}}\int_{0}^{\infty }u^{3}du\int_{-\infty }^{\infty }\cos
\left( us\right) \cos \left( \omega _{0}s\right) ds \\
&=&\frac{
\rlap{\protect\rule[1.1ex]{.325em}{.1ex}}h%
\tau }{\pi mR^{2}}\int_{0}^{\infty }u^{3}du\pi \left[ \delta \left( \omega
_{0}+u\right) +\delta \left( \omega _{0}-u\right) \right] =\frac{%
\rlap{\protect\rule[1.1ex]{.325em}{.1ex}}h%
\tau }{mR^{2}}\omega _{0}^{3}.
\end{eqnarray*}

From the diffusion constant and the damping it is possible to obtain the
following Fokker-Planck equation for the probability density, $\rho \left(
\omega _{0}\right) ,$ of frequencies of the rotor 
\[
\frac{\partial \rho }{\partial t}=\frac{\partial }{\partial \omega _{0}}%
\left( \tau \omega _{0}^{3}\rho \right) +\frac{1}{2}\frac{\partial ^{2}}{%
\partial \omega _{0}^{2}}\left( \frac{
\rlap{\protect\rule[1.1ex]{.325em}{.1ex}}h%
\tau }{mR^{2}}\omega _{0}^{3}\rho \right) , 
\]
whose (regular) stationary solution is 
\[
\rho =\frac{2mR^{2}}{
\rlap{\protect\rule[1.1ex]{.325em}{.1ex}}h%
}\exp \left( -\frac{2mR^{2}\omega _{0}}{
\rlap{\protect\rule[1.1ex]{.325em}{.1ex}}h%
}\right) . 
\]

A similar method may be used for the three-dimensional rotor\cite{Boyerrotor}
and the result is 
\begin{equation}
\rho =\left( \frac{2mR^{2}}{
\rlap{\protect\rule[1.1ex]{.325em}{.1ex}}h%
}\right) ^{2}\omega _{0}\exp \left( -\frac{2mR^{2}\omega _{0}}{%
\rlap{\protect\rule[1.1ex]{.325em}{.1ex}}h%
}\right) .  \label{3drotor}
\end{equation}

\subsection{Comparison between SED and QM}

The predictions of SED for the rigid rotor disagree with those of QM at
least in four respects, that will be illustrated in the following for the
particular case of the three-dimensional rigid rotor:

1. \emph{The distribution of positions or momenta in the minimal energy
state. }In quantum mechanics the eigenstates of the angular momentum squared
and the Hamiltonian of the rotor are, respectively, 
\begin{equation}
\mathbf{L}^{2}=
\rlap{\protect\rule[1.1ex]{.325em}{.1ex}}h%
^{2}l(l+1),E_{l}=\frac{
\rlap{\protect\rule[1.1ex]{.325em}{.1ex}}h%
^{2}}{2I}l(l+1),l=0,1,2...  \label{qrotor}
\end{equation}
so that the ground state corresponds to $\mathbf{L}^{2}=E=0.$ In contrast
the stationary solution in SED is given by eq.$\left( \ref{3drotor}\right) $
where there is a spherical distribution of angular momenta given by 
\begin{equation}
W(L)LdL=\frac{4}{
\rlap{\protect\rule[1.1ex]{.325em}{.1ex}}h%
^{2}}\exp \left( -\frac{2L}{
\rlap{\protect\rule[1.1ex]{.325em}{.1ex}}h%
}\right) LdL.  \label{srotor}
\end{equation}

2. \emph{The set of states}. As in the oscillator studied in section 3.3 the
set of possible states is quite different in QM and SED.

3. \emph{The spectrum}$.$ In QM the spectrum consists of the set of
frequencies 
\begin{equation}
\omega _{lj}=\left( 
\rlap{\protect\rule[1.1ex]{.325em}{.1ex}}h%
/2I\right) \left[ j(j+1)-l(l+1)\right] \rightarrow \left( 
\rlap{\protect\rule[1.1ex]{.325em}{.1ex}}h%
/I\right) \left( l+1\right) ,  \label{freqrotor}
\end{equation}
the latter frequencies corresponding to the transitions allowed in the
atomic dipole approximation. In sharp contrast SED predicts a continuous
spectrum, although the most intense absorption from the stationary state eq.$%
\left( \ref{qrotor}\right) $ corresponds to the maximum absorption, that may
be shown to be $\omega =
\rlap{\protect\rule[1.1ex]{.325em}{.1ex}}h%
/I$ \cite{dice}, in agreement with the QM result for the transition from the
ground to the first excited state, see eq.$\left( \ref{freqrotor}\right) .$
However QM predicts a sharp frequency whilst the SED prediction corresponds
to a wide band. In experiments the frequency is not sharp, but it is less
wide than the SED prediction. The disagreement between the QM prediction and
experiments is usually explained because the rigid rotator is not a
realistic model of a molecule. For instance molecules are not completely
rigid. The disagreement with experiments is greater in SED and it cannot be
explained as easily as in QM.

4. \emph{The specific heat. }There is also a discrepancy as shown in the
comparison between quantum and SED treatments \cite{Boyerrotor}. This point
will not be discussed here.

\subsection{A difficulty with the angular momentum}

The disagreement between the quantum prediction, eq.$\left( \ref{qrotor}%
\right) ,$ and the SED prediction, eq.$\left( \ref{srotor}\right) ,$ for the
rigid rotor is actually general and it puts a problem for any realistic
model of the rotation in quantum physics. For instance if we want to get a
picture of a rotating molecule. The quantum ground state of the rigid rotor
possesses \emph{zero angular momentum} and \emph{spherical symmetry, }but
these two properties are contradictory for any realistic interpretation. For
the sake of clarity let us consider a more realistic example, for instance
the molecule of carbon oxide, $CO,$ which may be modelled by a
three-dimensional rigid rotor. It consists of an oxigen atom and a carbon
atom at a distance which is very well known empirically. The ground state of
this molecule possesses zero angular momentum and \emph{therefore}
(according to the quantum formalism) spherical symmetry. Discarding
explanations which are bizarre for any realistic interpretation, like saying
that ``the form of the molecule emerges during the act of measurement'', the
meaning of spherical symmetry is unclear. The unique meaning compatible with
a physical picture is that the molecule is rotating randomly in such a way
that the probability distribution of the orientations of the axis in space
possesses spherical symmetry. However this is in conflict with the quantum
prediction that the total angular momentum is zero, \emph{dispersion-free}.
Therefore the mean square angular momentum is also zero. The situation is
quite common, it involves many molecules, atoms or nuclei. It seems that
either the standard quantum prediction is wrong (e. g. the ground state is
not physically achievable) or a \emph{realistic} physical model is not
possible.

A possible solution to the dilemma is that the quantum formalism actually
provides the total angular momentum of the molecule \textit{plus} the vacuum
fields that interact with it. If the ground state correspond to an
equilibrium of the system (e. g. the molecule) with the vacuum fields it is
plausible to assume that there is a continuous exchange of angular momentum
between the system and the vacuum fields so that the total angular momentum
(a conserved quantity) remains always zero. This is the case in the SED
treatment of the planar rigid rotor of the previous section. In fact, eq.$%
\left( \ref{rotor}\right) $ may be interpreted as the equation for the
balance of angular momentum. In fact the equation may be rewritten as the Z
component of the angular momentum vector, that is 
\[
\frac{d}{dt}(I\omega )=-\tau I\omega ^{3}+\tau I\stackrel{..}{\omega }%
+e\left( \mathbf{R}\times \mathbf{E}\right) _{z}, 
\]
where the change of the rotor angular momentum equals the radiated momentum
(the first two terms) minus the momentum absorbed from the ZPF. In summary
there is no real contradiction between the fact that the SED predicts a
distribution of angular momenta of \emph{the rotor alone} and our
interpretation of the QM prediction that the angular momentum of \emph{rotor
plus field} is stricly zero. I do not believe that SED is the correct
reinterpretation of QM, but I think that it illustrates adequately the
possible solution to the problem.

A similar solution may be given to the strange, if not paradoxical, quantum
prediction that a charged particle in a homogeneous magnetic field has zero
component of the angular momentum in the direction of the field, but the
energy is precisely the product of the field times the Bohr magneton. The
SED results are more intuitive, namely the energy of the equilibrium state,
eq.$\left( \ref{7.0}\right) ,$ agrees with the quantum ground energy, but
there is a component of the angular momentum in the direction of the field,
see eq.$\left( \ref{7.1a}\right) .$

\subsection{The hydrogen atom}

The hydrogen atoms is the most relevant nonlinear system within elementary
quantum mechanics, therefore a crucial test for the validity of SED. Once
the stationary state of the harmonic oscillator had been solved with
success, several authors devoted a big effort during the 1960's to study the
hydrogen atom in SED. Several approximation methods were proposed for
calculating the stationary state of the atom (modelled as two particles with
opposite charge, one of them at rest). The most successfull method devised
for the study of a charged particle in a potential well rests upon the
assumption that the classical constants of the motion change slowly. That is
the motion is close to the classical one, the action of the ZPF and the
radiation reaction giving rise to a slow diffusion in the space of classical
orbits. As every classical orbit is determined by the initial position and
velocity, $\left\{ \mathbf{r}_{0},\mathbf{v}_{0}\right\} ,$ the final result
of the calculation is a probability distribution in the phase space of
positions and velocities, $\left\{ \mathbf{r},\mathbf{v}\right\} ,$ that is
the same as the distribution of initial positions and velocities, $\left\{ 
\mathbf{r}_{0},\mathbf{v}_{0}\right\} ,$ if the state is stationary. This is
similar to what happens in the planar rigid rotator.

In the case of the hydrogen atom the result of the calculation did not
provide a stationary solution. In fact the prediction was that the atom is
not stable but ionizes spontaneously due to the orbits passing close to the
nucleus\cite{Claverie}.

That work has been criticized because such orbits cannot be treated with a
non-relativistic approximation, and a relativistic treatment could produce
an important change in the results. Actually the prediction of spontaneous
ionization made by SED analytical calculation is not a too strong argument
against the SED prediction. In fact the result depends crucially on the
electron orbits passing close to the nucleus, that would requiere a
relativistic treatment. Also quantum theory predicts that the free atom is
unstable against ionization at any finite temperature, no matter how small.
This trivially follows from the fact that the quantum partition function is
divergent, that is 
\[
Z=\sum_{n=1}^{\infty }\sum_{l=0}^{n-1}(2l+1)\exp \left( -\frac{E_{0}}{n^{2}}%
\right) \rightarrow \infty . 
\]
Therefore it is not too relevant if an approximation method used in SED has
an effect (spontaneous ionization) similar to the effect of a thermal
radiation in QM.

Furthermore numerical solutions of the hydrogen atom in SED have been made%
\cite{Cole} since 2003 that explain the stability of the atom. They led to
stationary distribution fairly close to the quantum prediction for the
position distribution in the ground sate. However more powerful calculations
made in 2015\cite{Nieu} predict a ionization of the atom. Numerical
calculations have the advantage that do not require approximations in the
differential equations, like the neglect of the dependence on position of
the ZPF (the electric dipole approximation). However the numerical methods
have uncertainties that may explain the discrepancy as commented above for
the early analytical treatment. See also \cite{Cole18}.

\subsection{Thermal equilibrium between radiation and matter. SED derivation
of Planck Law.}

Several authors have claimed that Planck's law may be derived from classical
postulates, usually within the framework of SED\cite{dice2},\cite{Boyer12}.
A derivation of the thermal radiation should follow from the study of the
thermal equilibrium between radiation and matter. In the framework of
standard quantum theory it leads to Planck's law, but here we are
considering the question whether it may be obtained from classical
electrodynamics. The difficulty is related to the fact that the equilibrium
radiation-matter should involve nonlinear systems. In particular the study
of equilibrium requires a balance between absorption of energy from the
radiation at a frequency and emission at a different frequency. Only in
these conditions it is possible to study the distribution of energy amongst
the different frequencies that is the essential purpose of a radiation law.
If we deal only with linear (harmonic) oscillators both the absorption and
emission of radiation take place at the same frequency.

The problem of thermal equilibrium was extensively studied in the first
decades of the 20th century and the conclusion was uncontroversial in my
opinion: If one assumes classical dynamics then thermal equilibrium is
achieved when the particles have the Maxwell-Boltzmann distribution and the
radiation the Rayleigh-Jeans spectrum\cite{vanVleck}. Thus there is a
contradiction between the derivation reported by van Vleck and the
derivations claiming that the classical equilibrium spectrum is given by
Planck\'{}s law. It was suggested that early derivations\cite{vanVleck}
involved Newtonian dynamics and that a study with relativistic dynamics
might led to Planck\'{}s law. However it has been shown that thermal
equilibrium of relativistic particles also leads to the Rayleigh-Jeans law%
\cite{Blanco}. A different question is whether Planck law may be derived for
systems of charged particles immersed in the ZPF field plus additional
radiation and we assume thermal equilibrium of that radiation with the
particles. In these conditions Planck spectrum is obtained\cite{Boyer12}, 
\cite{Boyer18}.

A related result is the classical derivation of the Davies-Unruh effect
initially derived from quantum electrodynamics\cite{Davies}, \cite{Unruh}.
It is interpreted in quantum theory as the production of photons with Planck
distribution of frequencies when a detector moves in the vacuum with
accelerated motion. The result may be got in SED with the interpretation
that the spectrum of the ZPF appears as thermal when seen from an
accelerated reference frame\cite{Boyer13}, \cite{Milonni}.

\section{SED as a clue for a realistic interpretation of quantum mechanics}

We have seen that calculations of several linear systems within SED provide
a remarkable agreement with the predictions of QM. On the other hand the
realistic interpretation of SED is rather obvious. Thus the question arises,
offers SED the realistic interpretation of QM which we are searching for?.
Unfortunately the answer is in the negative, the difficulties of SED for the
interpretation of phenomena associated to nonlinear systems seem
unsourmontable.

I propose that geting a realistic interpretation of QM would be possible
accepting the general ideas of SED but rejecting many of the particular
assumptions. The general ideas to be retained are the following: 1) Nuclei,
atoms or molecules (but maybe not elementary particles like electrons) are
bodies with well defined size and form following definite, but highly
irregular, trajectories. (If the bodies are composite, like atoms, they may
suffer deformations). 2) The motion is strongly influenced by the
fluctuations of the vacuum fields.

A summary of the clues provided by SED for a realistic interpretation of
(non-relativistic) quantum mechanics follows.

\textit{The attempt to interpret the quantum mechanics of particles alone is
misleading if quantum fields, in particular vacuum fields, are not included. 
}

\textit{The quantum ground state of a particle in a potential well
corresponds to a stationary state of the particle performing a highly
irregular (stochastic) motion driven by vacuum fields. There is a dynamical
equilibrium between absorption from and emission of radiation to the vacuum
fields. The interaction gives rise to probability distributions of
coordinates and momenta of the particle that agree with quantum predictions
for linear systems, but for nonlinear ones there is disagreement. Radiative
corrections (e, g. Lamb shift) have a transparent interpretation. }

\textit{The study of coupled oscillators at zero Kelvin provides an
intuitive picture of entanglement as a correlation between quantum
fluctuations mediated by the vacuum fiels. At a finite temperature it gives
a simple realistic interpretation of the Debye theory of specific heats of
solids.}

\textit{\ The motion of particles is highly irregular due to the interaction
with the vacuum fields. In particular the free particle possesses a
conserved canonical momentum with an associated inertial motion but,
superimposed to this, it has a random motion with high velocity that cannot
be studied adequately in the nonrelativistic approximation. This derives
from the spectrum,} $S_{E}\left( \omega \right) \varpropto \omega ^{3},$%
\textit{\ of the vacuum radiation fields: At short times the motion is
governed by the high frequencies where} $S_{E}\left( \omega \right) $ 
\textit{is large thus inducing a rapid erratic motion, at long time it is
governed by the low frequencies where} $S_{E}\left( \omega \right) $ \textit{%
is small, whence the memory of the initial velocity is lost slowly. This
behaviour is very different from Brownian motion.}

\textit{The spectrum of the radiation absorbed or emitted by a particle in a
potential well badly fails to reproduce the quantum spectrum, except in the
trivial case of the harmonic oscillator whose spectrum consists of a single
frequency. This fact suggests that the back action of the particles on the
vacuum radiation field and/or the inclusion of many fields would be
essential for the prediction of spectra, e. g. of atoms.}

\textit{The stochastic commutator provides a hint for a realistic
interpretation of the quantum commutation rules as a disguised form of
stablishing the properties of some peculiar stochastic processes. The main
peculiarity is the fact that the spectra of the processes is usually odd
with respect to a change }$\omega \rightarrow -\omega $ \textit{of the
frequency}. \textit{Thus I propose that the reason for the success of
formulating QM with noncommuting mathematical objects is the fact that the
basic stochastic processes involved have spectra that are odd with respect
to the change }$\omega \rightarrow -\omega .$\textit{\ }

\textit{The SED calculation of the specific heat of solids provides an
argument for the continuity of the energies of ions. Quantized oscillations
of the ions in a solid (phonons) are not particles, the energies of the
normal modes of the set of ions having a continuous distribution of
energies. The mean energy of a vibration mode of the ions is the same as the
mean energy of a radiation mode with the same frequency.}

\textit{The prediction that the angular momentum is zero, dispersion-free,
in some cases is possibly the most paradoxical prediction of QM. In fact a
molecule, or the electron in the ground state of a hydrogen atom, are
predicted to be in a state with spherical symetry. A realistic
interpretation is possible only if we assume that there is a random rotation
having zero angular momentun on the average, but the mean squared angular
momentum being diffeent from zero. Our study suggests a ralistic
interpretation assuming that the dispersion-free momentum refers to the
addition of two highly correlated random angular momenta, namely those of
the material system plus the vacuum fields, these modified by the presence
of matter.}

\end{document}